\documentclass[a4paper,11pt]{article}
\usepackage{jheppub} 

\usepackage[mathlines]{lineno}

\let\oldequation\equation
\let\oldendequation\endequation

\renewenvironment{equation}
  {\linenomathNonumbers\oldequation}
  {\oldendequation\endlinenomath}

\let\oldalign\align
\let\oldendalign\endalign

\renewenvironment{align}
  {\linenomathNonumbers\oldalign}
  {\oldendalign\endlinenomath}

\let\oldalignat\alignat
\let\oldendalignat\endalignat

\renewenvironment{alignat}
  {\linenomathNonumbers\oldalignat}
  {\oldendalignat\endlinenomath}

\usepackage[dvipsnames,table,xcdraw]{xcolor}    
\usepackage{mathtools}                          
\usepackage{float}                              
\usepackage{siunitx}                            

\usepackage{bm}

\usepackage{multirow,booktabs}

\newcommand{\diff}{{\rm d}}

\newcommand{\CR}{C_{\rm R}^{}}

\newcommand{\ptrel}{\boldsymbol{\kappa}}
\newcommand{\ptevol}{\boldsymbol{\kappa}_{*}}
\newcommand{\mass}{m}
\newcommand{\massALT}{m_{*}}
\newcommand{\invtf}{\tau^{-1}}
\newcommand{\invtfALT}{\tau^{-1}_{*}}
\newcommand{\invtfALTarg}[1]{\tau^{-1}_{*\,#1}}
\newcommand{\tform}{\tau}

\newcommand{\angleVAR}{\theta}
\newcommand{\angleVARALT}{\theta_{*}}

\preprint{MIT-CTP/5761}

\title{\boldmath Assessing Uncertainties in Parton Showers at Double Logarithmic Accuracy for Jet Quenching Studies}

\author[a,b]{Carlota Andres}
\author[b,c]{, Liliana Apolinário}
\author[d]{, N\'estor Armesto}
\author[b,c]{, Andr\'e Cordeiro}
\author[d]{, Fabio Dominguez}
\author[b,c]{, Jos\'e Guilherme Milhano}
\affiliation[a]{Center for Theoretical Physics, Massachusetts Institute of Technology, Cambridge, MA 02139, USA}
\affiliation[b]{LIP - Laborat\'{o}rio de Instrumentação e F\'{i}sica Experimental de Part\'{i}culas\\
Av. Prof. Gama Pinto, 2, 1649-003, Lisbon, Portugal}
\affiliation[c]{Departamento de Física, Instituto Superior Técnico, Universidade de Lisboa\\
Av. Rovisco Pais 1, 1049-001 Lisbon, Portugal}
\affiliation[d]{Instituto Galego de F\'{\i}sica de Altas Enerx\'{\i}as IGFAE,
Universidade de Santiago de Compostela,\\ 15782 Santiago de Compostela, Galicia-Spain}

\emailAdd{c\_andres@mit.edu}
\emailAdd{liliana@lip.pt}
\emailAdd{nestor.armesto@usc.es}
\emailAdd{andre.cordeiro@tecnico.ulisboa.pt}
\emailAdd{fabio.dominguez@usc.es}
\emailAdd{gmilhano@lip.pt} 

\abstract{We present a systematic study of how different choices of ordering and phase-space constraints in parton showers affect the space-time structure of vacuum parton cascades and their interface with jet quenching models. Using a simplified Monte Carlo shower implemented at double logarithmic accuracy, we analyse variations in emission patterns and resulting phase-space arising from three ordering variables: inverse formation time, invariant mass, and opening angle. These are coupled with two kinematic reconstruction schemes defined by different phase-space constraints. We show that, while global features are relatively stable, differences emerge in the temporal evolution of the cascade. To probe the impact of these differences, we introduce a simplified model for in-medium energy loss based on formation time and colour decoherence, enabling us to evaluate the sensitivity of quenching observables to the underlying space-time structure of the vacuum shower. We further quantify the role of time-ordering violations and propose strategies to preserve a consistent space-time interpretation. Lastly, we explore a range of alternative quenching models confirming the robustness of our conclusions. Our findings highlight the importance of maintaining a coherent space-time structure in parton shower algorithms when modelling jet propagation in an extended QCD medium, as this structure becomes a physically meaningful and testable component of the jet itself.}

\begin{document}
\maketitle
\flushbottom

\section{Introduction}
\label{sec:intro}

The simulation of the branching process of a parton in Quantum Chromodynamics (QCD), the parton shower, forms the foundation of Monte Carlo event generators used in high-energy particle collisions~\cite{Campbell:2022qmc}. Initially developed at double leading logarithmic (DLL) accuracy, parton showers factorize successive emissions, resulting in a Markovian process that captures the dominant soft and collinear divergences. Current advancements aim at reducing theoretical uncertainties by incorporating corrections beyond DLL accuracy~\cite{Bewick:2019rbu,Dasgupta:2020fwr,Forshaw:2020wrq,Nagy:2020rmk,Nagy:2020dvz,Bewick:2021nhc,vanBeekveld:2022ukn,Herren:2022jej,vanBeekveld:2023chs,FerrarioRavasio:2023kyg,vanBeekveld:2024wws}, which requires careful matching to exact higher-order matrix elements. While different evolution variables for the parton shower (such as virtuality, transverse momentum, or formation time) yield identical results at sufficiently high energies, deviations become prominent at lower energies, manifesting as single logarithmic or non-logarithmic contributions. Additionally, the method used for kinematic reconstruction of the parton momenta significantly impacts these differences, even when using the same evolution variable. Such implementation details, while part of the intrinsic uncertainty in vacuum showers, become particularly critical in the presence of a strongly interacting medium.

In heavy-ion collisions, the evolution of parton showers~\cite{Mehtar-Tani:2013pia, Blaizot:2015lma, Apolinario:2022vzg} occurs within a Quark-Gluon Plasma (QGP), whose dynamics are commonly modelled using relativistic hydrodynamics after initial collision times (in the range $0.5 - 1~\rm{fm/c}$). Due to this space-time description of the QGP, parton evolution variables which can be connected to a coordinate space (e.g., formation time) are preferable to momentum-based ones. However, current jet quenching models either consider medium effects only after a partially developed vacuum shower or incorporate them throughout the entire shower evolution using momentum-based variables with assigned space-time mappings~~\cite{Lokhtin:2005px,Schenke:2009gb,Deng:2010mv,Casalderrey-Solana:2014bpa,Caucal:2018dla,Luo:2023nsi,Armesto:2009fj,Zapp:2013vla,Putschke:2019yrg}. As a result, the uncertainties inherent in the vacuum shower translate into significant ambiguities in the space-time picture, affecting energy-loss predictions crucial for jet quenching studies.

This work systematically assesses how variations in evolution variables and kinematic reconstruction constraints within parton showers influence energy-loss calculations in jet quenching scenarios. We quantify these uncertainties by constructing a simplified Monte Carlo model for a massless quark-initiated cascade, explicitly varying evolution schemes and kinematic constraints\footnote{Some early considerations along these lines were explored in~\cite{Takacs:2020tqn}.
}.
Our analysis focuses on how these technical choices alter observable distributions, Lund plane structures~\cite{Dreyer:2018nbf,Andrews:2018jcm,Apolinario:2024equ}, and critically, the occurrence of formation-time inversions -- events violating the expected causal structure of medium evolution. Understanding these aspects is essential since formation-time inversions, negligible in vacuum due to their minimal impact on DLL accuracy, become highly relevant in a medium, potentially disrupting the causal interface between the shower evolution and the medium dynamics. Thus, by clearly delineating and quantifying these uncertainties, we aim to guide future developments in jet quenching modelling towards more robust and physically consistent implementations.

The contents of this paper are organized as follows. In section~\ref{sec:toymc}, we introduce the toy Monte Carlo parton shower setup that is used as our vacuum baseline. Details regarding the ordering variable choices (formation time, mass, and opening angle), kinematic reconstruction constraints (subsection~\ref{sec:toymc-ordering}) and the algorithm (subsection~\ref{sec:toymc-algo}) to generate the emission pattern are present in this section. It is followed (subsection~\ref{sec:toymc-plots}) by a discussion on how the uncertainties arising from variations in these settings are expressed in terms of the radiation phase-space, expressed in terms of the relative transverse momentum and opening angle of successive parton splittings.
In section~\ref{sec:energyloss}, we present a simplified quenching prescription to account for in-medium effects. Once this model is introduced (subsection~\ref{sec:quenching-model}), an in-depth study of how jet survival probabilities depend on the space-time structure inherited from the vacuum shower follows in subsection~\ref{sec:energyloss-hi}. This includes an analysis of the quenched phase-space, the role of formation time violations and their mitigation, as well as the exploration of alternative quenching prescriptions. We conclude in section~\ref{sec:conclusions}, where we summarize our findings and outline implications for future theoretical and phenomenological developments.

\section{Toy Monte Carlo Shower: Vacuum Baseline}
\label{sec:toymc}

Our goal is to develop a simplified Monte Carlo parton shower framework capable of systematically varying the ordering variables while maintaining a consistent starting and stopping criterion across different implementations. This allows us to isolate and assess the effects solely due to the choice of ordering variables and reconstruction schemes on parton shower characteristics and jet quenching observables. 

A parton shower models multiple sequential parton emissions, typically formulated as a Markovian branching process. At leading logarithmic accuracy, the probability of a specific splitting $a \rightarrow b + c$ occurring at a resolution scale $s$ with energy fraction $z$ of parton $b$ is given by
\begin{equation}
\diff \mathcal{P}_{a \rightarrow bc} = \frac{\diff s}{s} \frac{\alpha}{2\pi} \hat{P}_{a \rightarrow bc}(z) \diff z\,,
\label{eq:prob-split}
\end{equation}
where $\alpha$ is the QCD coupling constant and $\hat{P}_{a \rightarrow bc}(z)$ are, for the present work, the unregularized leading-order Altarelli-Parisi splitting kernels~\cite{Gribov:1972ri,Gribov:1972rt,Altarelli:1977zs,Dokshitzer:1977sg}. The survival probability for no emission between scales $s_{\rm prev}$ and $s_{\rm next}$ is expressed by the Sudakov form factor
\begin{equation}
\Delta_{a}(s_{\rm prev}, s_{\rm next}) = \exp \left\{- \int^{s_{\rm prev}}_{s_{\rm next}} \frac{\diff s}{s} \frac{\alpha(s)}{2\pi} \int_{\Gamma(s)} \hat{P}_{a}(z) \diff z \right\}\,,
\label{eq:sudakov}
\end{equation}
where $\Gamma(s)$ defines the allowed emission phase-space. 

In our setup, we adopt the double leading logarithmic (DLL) approximation, where only the singular parts of the leading-order splitting kernels are retained, and the QCD coupling $\alpha$ is taken to be fixed. The phase-space $\Gamma(s)$ depends explicitly on the ordering variable and thus determines the scale-dependent constraints on the allowed energy fractions. These will be made explicit in the following sections when defining the starting and stopping conditions for each ordering scheme. Finally, we note that while this framework is general and can accommodate the full evolution of a partonic cascade including gluon-initiated branches and flavour-changing processes, our focus here is on only emissions off a quark along its primary branch. This choice enables a cleaner evaluation of how the ordering variable and kinematic scheme affect the resulting space-time picture, a feature most relevant in the context of jet quenching in a QCD medium. Accordingly, at the accuracy we work with, we restrict ourselves to a sequence of emissions $a\ (\text{quark}) \rightarrow b\ (\text{gluon}) + c\ (\text{quark})$. Despite this simplification, our setup can be readily extended to include other channels if needed.

\subsection{Ordering Variables and Kinematic Reconstruction}
\label{sec:toymc-ordering}

To systematically study the influence of ordering choices, we introduce three variables commonly used in parton shower algorithms: the inverse formation time\footnote{While not exactly the formation time as we will employ, time-ordered parton showers based on light-cone formation time have been previously explored in~\cite{Nagy:2009vg,Nagy:2014nqa,vanBeekveld:2022zhl}.} $\invtf$, the invariant mass squared $\mass^2$ and the squared opening angle $\angleVAR^2$. These variables reflect different physical interpretations of the scale governing the branching process. Formation time can be connected to the space-time picture of parton splittings, the invariant mass governs the virtuality of the intermediate states, and the opening angle is tied to the angular structure of QCD radiation.

We use light-cone coordinates\footnote{We adopt the convention $p^\pm = (p^0 \pm p^3)/\sqrt{2}$ and $\boldsymbol{p} = (p^1, p^2)$ for the transverse components.} to describe the momenta of the partons. In this basis, the energy of the parent quark $a$ is approximated as $E = p_a^+/\sqrt{2}$, and the energy fraction of the emitted gluon $b$ is given by $z = p_b^+/p_a^+$. In the following, we explore two reconstruction schemes that yield different proxy definitions for the ordering variables:

\paragraph{Momentum Scheme:} In this scheme, the relevant variables are reconstructed from the relative transverse momentum between the two daughter partons, $b$ and $c$. Denoting this quantity as $\ptrel = (1-z)\,\boldsymbol{p}_b - z\,\boldsymbol{p}_c$, and requiring energy-momentum conservation in the $a \rightarrow b + c$ splitting, one can write the chosen ordering variables as
\begin{equation}
\invtf = \frac{|\ptrel|^2}{E z(1-z)}\,,\ \ \ \ 
\mass^2 = \frac{|\ptrel|^2}{z(1-z)}\,,\ \ \ \ 
\angleVAR^2 = \frac{|\ptrel|^2}{[E z(1-z)]^2}\, .
\label{eq:variables-momentum-scheme}
\end{equation}
This formulation is advantageous because the transverse momentum cutoff $|\ptrel| \geq \Lambda$ can be directly imposed, ensuring emissions lie in the perturbative regime.

\paragraph{Mass Scheme:} Alternatively, one may define the ordering variables using the invariant mass of the parent parton $p_a^2$, while still computing $|\ptrel|$ via momentum conservation. This leads to:
\begin{equation}
\invtfALT = \frac{p_a^2}{E}\,,\ \ \ 
\massALT^2 = p_a^2\,,\ \ \ \ 
\angleVARALT^2 = \frac{p_a^2}{E^2 z(1-z)}\, .
\label{eq:variables-mass-scheme}
\end{equation}
This scheme has the advantage of providing a more direct mapping between the evolution scale and the virtuality of the splitting parton, simplifying the treatment of energy-momentum conservation in some implementations. 
\paragraph{}

Both schemes are implemented using a veto algorithm that ensures consistency with the specified ordering and kinematic criteria. The differences between the schemes, arising from their treatment of kinematics and the resulting emission phase-space, will be central to our analysis of shower-induced uncertainties.

Finally, we stress that the term kinematic scheme (used interchangeably with reconstruction scheme throughout the text) refers strictly to the set of phase-space constraints imposed on the parton branching process. It does not involve any momentum reshuffling, recoil prescription, or subtraction procedure such as those used in~\cite{Bewick:2019rbu,Dasgupta:2020fwr,vanBeekveld:2022ukn}. Our analysis is limited to the evolution of kinematic variables directly obtained from the parton shower algorithm, allowing us to isolate the impact of ordering and phase-space structure without extra considerations from four-momentum reconstruction.

\subsection{Parton Shower Algorithm}
\label{sec:toymc-algo}

Having defined the ordering variables and reconstruction schemes, we now describe the algorithm used to build a vacuum parton shower (specifically, the gluon-emission pattern off a quark) under the  DLL approximation. The shower is evolved through successive emissions by sampling the Sudakov form factor, with emission probabilities constrained by both a minimum and maximum scale for each ordering variable. These bounds, referred to as \emph{starting} and \emph{stopping} conditions, define the valid phase-space for each emission and are implemented in a scheme-consistent manner.

\paragraph{\textit{Stopping} conditions:} We require all emissions to lie within a perturbative regime defined by a lower cutoff $\Lambda$, which is imposed on the relative transverse momentum $|\ptrel|$. In the momentum scheme, the transverse momentum $|\ptrel|$ can be computed directly from the ordering variable $s$ through a one-to-one relation $|\ptrel| = \kappa(s, z)$, allowing the condition $|\ptrel| \geq \Lambda$ to be imposed explicitly. In contrast, the mass scheme lacks a direct mapping from $s$ to $|\ptrel|$, as the later depends on the daughter momenta:
\begin{equation}
|\ptrel|^2 = z(1-z)p_a^2 - (1-z)p_b^2 - zp_c^2\,.
\label{eq:relative-pt}
\end{equation}

To overcome this, we define a proxy variable $|\ptevol|^2 = z(1-z) p_a^2$ and require $|\ptevol| > \Lambda$. Since this condition alone does not guarantee $|\ptrel| \geq \Lambda$, the invariant masses of both daughters ($b$ and $c$) must be determined in parallel, starting from $p_b^2 < z p_a^2$ and $p_c^2 < (1-z) p_a^2$, and resampling both splittings whenever the condition $|\ptrel| \geq \Lambda$ is violated\footnote{
The condition $|\ptrel| \geq \Lambda$ implies $\invtfALTarg{a} > \invtfALTarg{b} + \invtfALTarg{c}$, yielding a stronger time ordering constraint than in the momentum scheme. It also implies strictly decreasing invariant masses, $p^2_a > p^2_{b,c}$. As a direct consequence of these facts, time and mass ordered showers become equivalent in this scheme.
}. This provides a conservative yet implementable stopping condition without requiring a complete kinematic reconstruction prior to vetoing. Consequently, in this scheme one must always try for a $g\to gg$ splitting to determine the $\ptrel$ of a quark branch splitting. Although alternative methods exist \cite{Webber:1986mc}, such as randomly selecting which daughter to sample first or always choosing the more energetic parton first, these differences are expected to be subdominant in the double logarithmic approximation.

\paragraph{\textit{Starting} conditions:} The maximum scale $s_{\text{max}}$ defines the upper bound for the first emission. A consistent choice across all ordering variables is to impose $\tau^{-1} \leq E$, with $E$ the energy of the initiating quark. This condition ensures that the shower evolution is well separated from the hard scattering scale, preserving a physically motivated time hierarchy. Special care is needed when using the angle as an ordering variable. In the momentum scheme, the emission angle satisfies $\theta^2 = \tau^{-1} / [E z(1-z)] \leq 1/(z(1-z))$. %
Since $z(1-z) \leq 1/4$, we choose $\angleVAR^2 \leq 4$ as a natural upper bound.
In the mass scheme, imposing $\angleVARALT^2 \leq 4$ automatically ensures\footnote{This condition also guarantees the parton shower fills the entire phase-space $\Lambda \leq |\ptrel| < E/2$ while correctly restricting $z$ with the soft-regulator.} $\angleVAR^2 \leq 4$.
\paragraph{}

The procedures described above allow for a consistent definition of the boundaries of the emission phase-space $\Gamma(s)$ across ordering variables. Taking the inverse formation time in the momentum scheme as an example, we impose $\invtf \leq E$ and require $|\ptrel|^2 \geq \Lambda^2$, which implies $z(1-z) \geq \Lambda^2/(E \invtf)$. This fully determines the available phase-space as $z \in [z_{\rm min}, 1 - z_{\rm min}]$, where
\begin{equation}
z_{\rm min} = \frac{1}{2} - \frac{1}{2} \sqrt{1 - \left( \frac{s_{\rm min}}{s} \right)^b }\,,
\label{eq:zmin}
\end{equation}
and $s_{\rm min} = 4 \Lambda^2 / E$ with $b = 1$. By a similar reasoning, the parameters $s_{\rm min}$, $s_{\rm max}$ and $b$ for other ordering variables are obtained and summarized in Table~\ref{tab:dla-sudakov-parameters}. Although this results in a potentially complex integral in Eq.~\eqref{eq:sudakov}, this can be handled using a veto algorithm~\cite{Bierlich:2022pfr,Lonnblad:2012hz}: the survival probability is computed in an extended phase-space $\bar{\Gamma}(s)$, and emissions outside the true phase-space $\Gamma(s)$ are rejected. The scale of the rejected emission becomes the new upper limit for further trial emissions. This process continues until a resolvable splitting is found or the trial scale drops below $s_{\min}$. The extended phase-space can be identified by defining
\begin{equation}
z_{\rm cut}(s) = \frac{1}{4} \left( \frac{s_{\rm min}}{s} \right)^b \, .
\end{equation}
The survival probability becomes
\begin{align}
\Delta_{\rm R} (s_{\rm prev}, s_{\rm next}) &= \exp \left\{ - \frac{\alpha\CR}{\pi} \int^{s_{\rm prev}}_{s_{\rm next}} \frac{\diff s}{s} \int^{1}_{z_{\rm cut}(s)} \frac{\diff z}{z} \right\} \nonumber \\ 
&= \exp\left\{ - \frac{\alpha \CR}{\pi} \, \frac{b}{2} \left[ \ln^2\left( \frac{4^{1/b} s_{\rm prev}}{s_{\rm min}} \right) - \ln^2\left( \frac{4^{1/b} s_{\rm next}}{s_{\rm min}} \right) \right] \right\},
\label{eq:sudakov-explicit}
\end{align}
highlighting the double-logarithmic enhancement characteristic of QCD radiation.

\begin{table}[h]
\centering
\begin{tabular}{c c c c}
\toprule
Ordering Variable $s$ & $s_{\min}$ & $s_{\max}$ & Regulator $b$ \\
\midrule
$\invtf$ & $4\Lambda^2/E$ & $E$ & 1 \\
$\mass^2$ & $4\Lambda^2$ & $E^2$ & 1 \\
$\angleVAR^2$ & $16\Lambda^2/E^2$ & $4$ & $1/2$ \\
\bottomrule
\end{tabular}
\caption{Starting ($s_\textrm{max}$) and stopping ($s_\textrm{min}$) scales for each ordering variable, along with the soft regulator exponent $b$ used in the Sudakov integral.}
\label{tab:dla-sudakov-parameters}
\end{table}

With these definitions, the parton shower proceeds through the following steps:
\begin{enumerate}
\item \textbf{Sample the ordering variable.} Draw a trial scale $s_{\rm trial}$ using the Sudakov form factor between $s_{\rm prev}$ and $s_{\rm next}$ by sampling a uniformly distributed random number $\mathcal{R}_s \in [0, 1]$ and solving:
\begin{equation}
\Delta_{\rm R}(s_{\rm prev}, s_{\rm trial}) = \mathcal{R}_s\,.
\end{equation}

\item \textbf{Sample the energy fraction.} Draw $z_{\rm trial}$ from the splitting kernel over the allowed region $[z_{\rm cut}(s), 1]$, by likewise sampling a random number $\mathcal{R}_z \in [0, 1]$ and solving:
\begin{equation}
\frac{\ln(z_{\rm trial} / z_{\rm cut})}{\ln(1 / z_{\rm cut})} = \mathcal{R}_z \, .
\end{equation}
\item \textbf{Check resolution conditions.} Compute kinematic proxies based on the chosen scheme:
\begin{itemize}
\item In the \emph{momentum scheme}, check $|\ptrel| \geq \Lambda$ and $\angleVAR^2 \leq 4$.
\item In the \emph{mass scheme}, try for $|\ptevol| > \Lambda$, $\angleVARALT^2 \leq 4$, and ensure $|\ptrel| \geq \Lambda$ for the previous splitting using the procedure explained above.
\end{itemize}
If the conditions are satisfied, accept the emission, create daughter partons with energy fractions $zE$ and $(1-z)E$, and iterate the procedure along the primary branch only (i.e., for the particle that retains energy fraction $(1-z)E$). Otherwise, veto the emission and retry from the new scale $s_{\rm prev} = s_{\rm trial}$.

\item \textbf{Terminate when unresolvable.} When the maximum allowed splitting scale is below $s_{\rm min}$, no further splittings can be resolved. As a result, the branching terminates and the parton is marked as \textit{final}.
\end{enumerate}

This procedure ensures that all sampled emissions satisfy physical constraints tied to the ordering variable, and allows for a controlled comparison across reconstruction schemes. Differences in the resulting parton cascades can thus be attributed directly to the definitions of ordering variable and kinematics. Importantly, we have verified that all shower implementations yield identical results in the double logarithmic limit, confirming the consistency of this framework at asymptotically large energies.

\subsection{Baseline Uncertainties}
\label{sec:toymc-plots}

We now assess how variations in the ordering variable and reconstruction scheme influence the structure of the vacuum parton shower, focusing on the resulting uncertainties. It is important to emphasize from the outset that these results are not intended to define physical observables, but rather to serve as internal consistency checks of our shower implementation and as a reference baseline for assessing potential medium-induced effects in later sections.

\begin{figure}[ht]
\centering
\includegraphics[width=.495\textwidth]{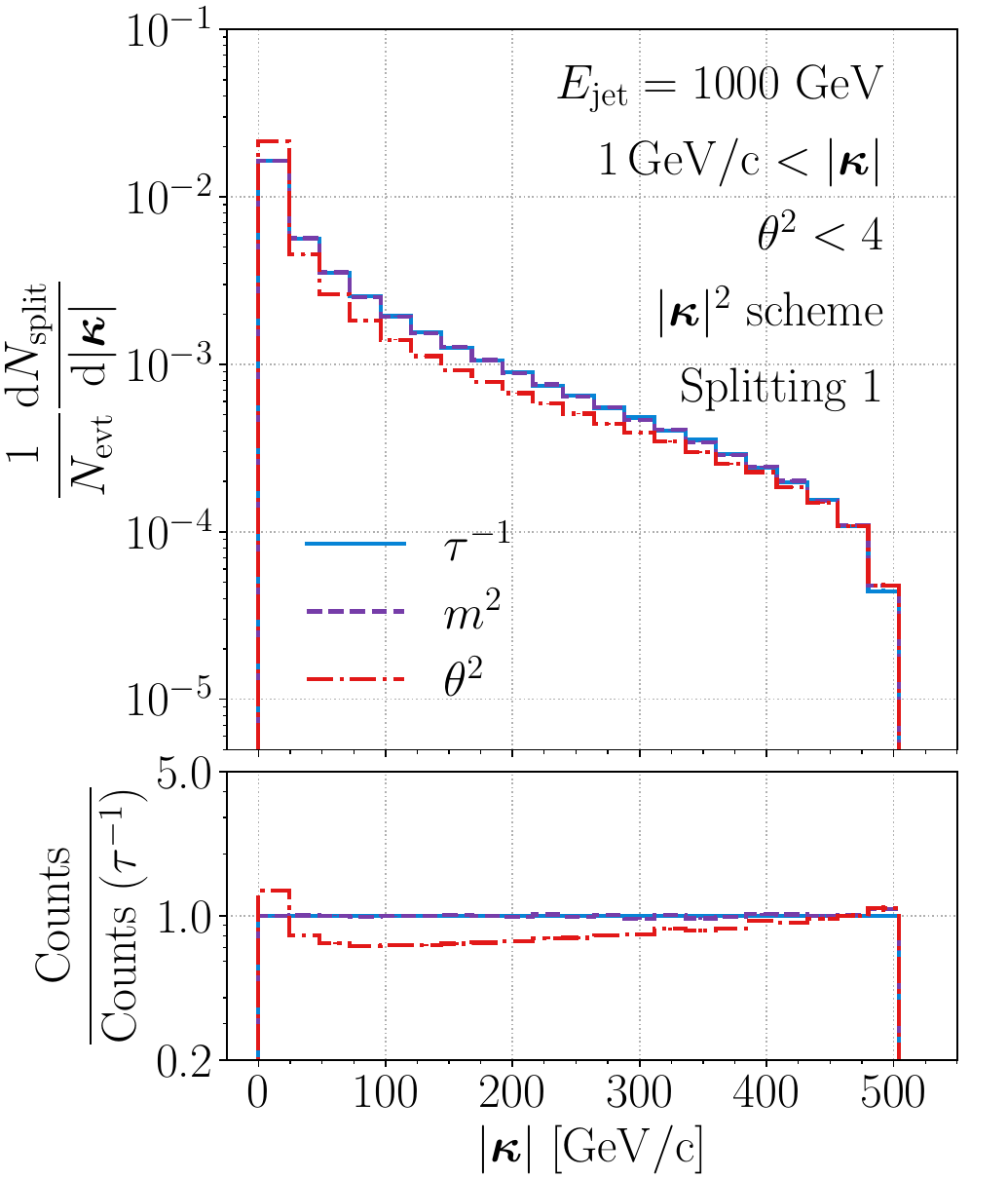}
\includegraphics[width=.495\textwidth]{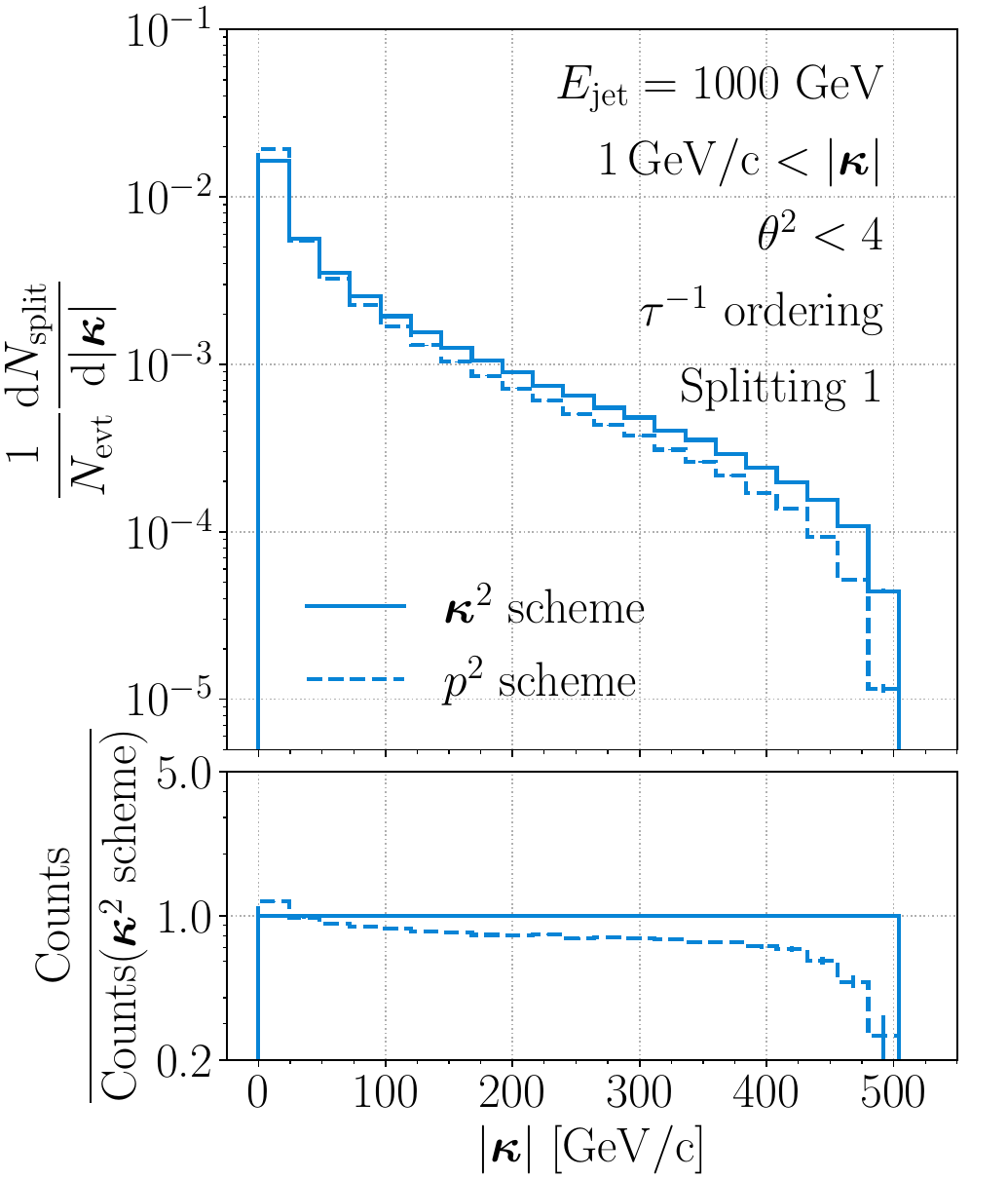}
\caption{Relative transverse momentum distribution of the first gluon emission off a quark for: \textbf{Left:} three different ordering variables, $\invtf$ (blue solid line), $\mass^2$ (purple dashed line), and $\angleVAR^2$ (red dash-dotted line); \textbf{Right:} two reconstruction schemes, momentum-scheme (blue solid line) and mass-scheme (blue dashed line). The bottom panels illustrate the ratios w.r.t a designated reference (formation time, in the left panel, momentum scheme, in the right panel)}
\label{fig:kt_1st}
\end{figure}

To begin with, we consider the relative transverse momentum of the first emission along the quark branch $|\ptrel|$. This helps us evaluate whether the starting conditions are consistently defined across the different parton shower algorithm choices. As shown in figure~\ref{fig:kt_1st}, all ordering variables lead to nearly identical distributions for this quantity, underscoring that the starting condition of the shower is consistently imposed across all configurations. This supports the conclusion that our procedure for defining the initial scale $s_{\text{max}}$ results in equivalent starting conditions for all ordering choices, as intended. The most noticeable differences arise from the choice of kinematic reconstruction scheme, though they are largely confined to the high-$\kappa$ region, where emissions are relatively rare.

\begin{figure}[ht]
\centering
\includegraphics[width=.495\textwidth]{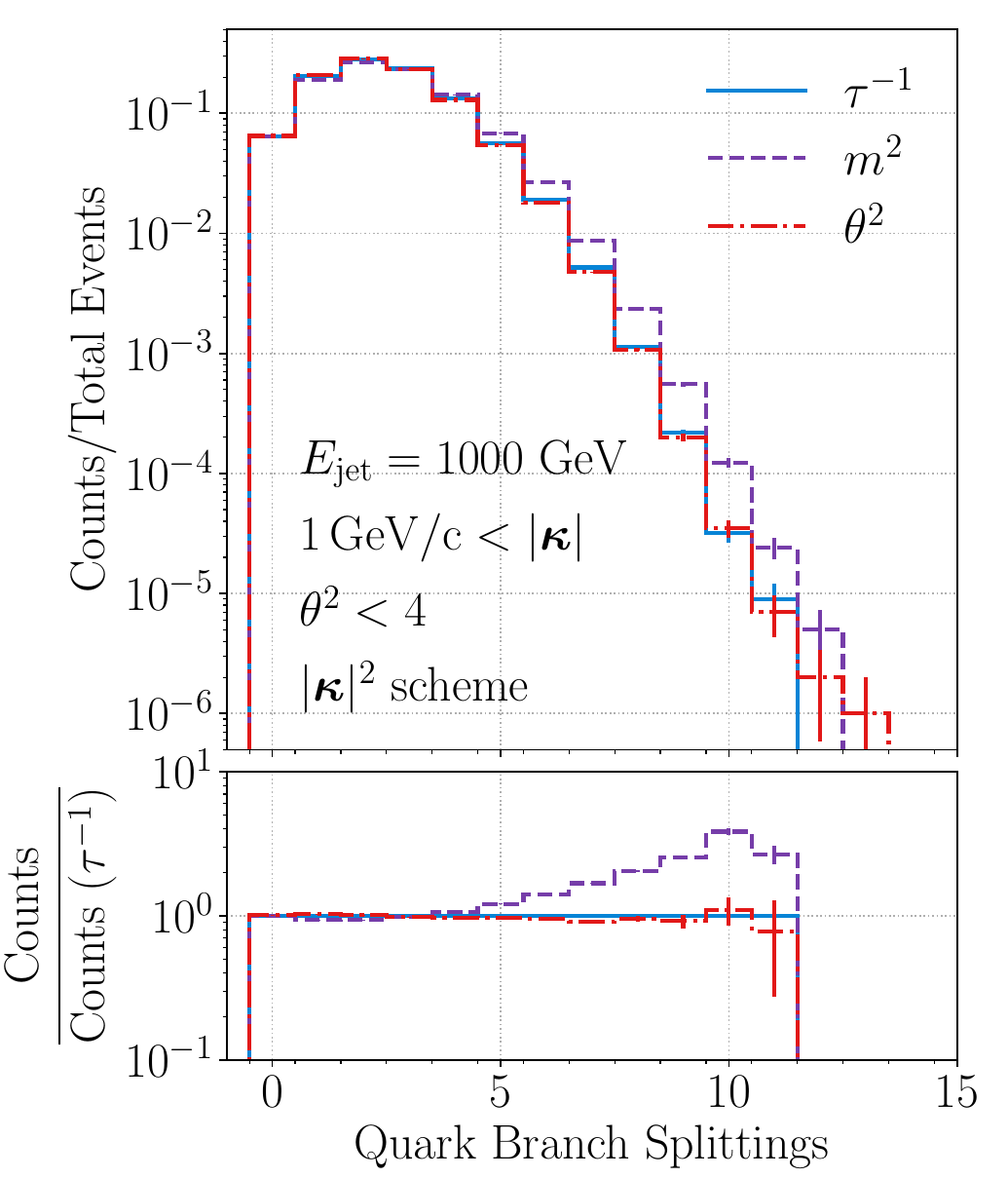}
\includegraphics[width=.495\textwidth]{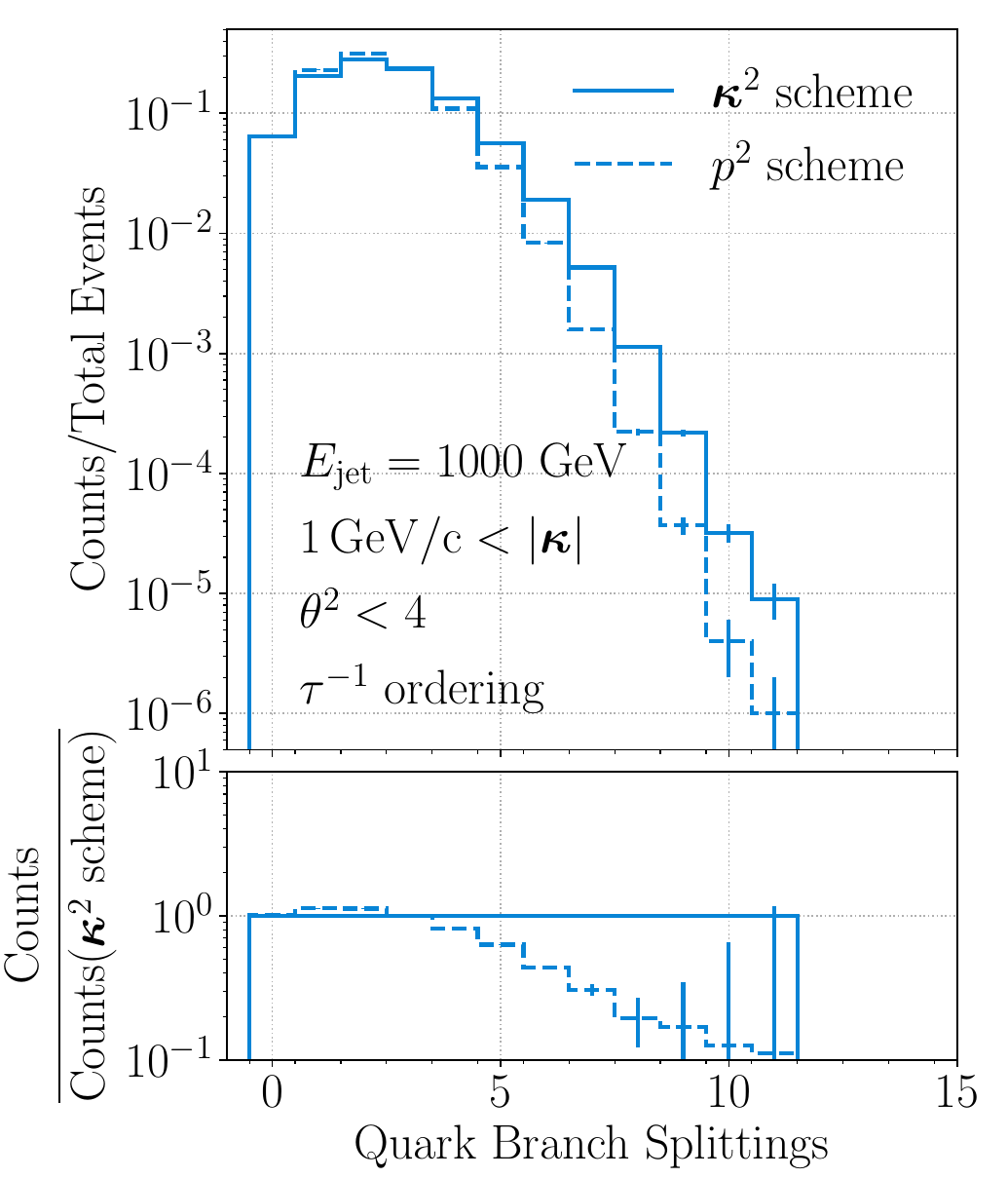}
\caption{Number of gluon emissions along the quark branch for: \textbf{Left:} three different ordering variables, $\invtf$ (blue solid line), $\mass^2$ (purple dashed line), and $\angleVAR^2$ (red dash-dotted line); \textbf{Right:} two reconstruction schemes, momentum-scheme (blue solid line) and mass-scheme (blue dashed line). The bottom panels illustrate the ratios w.r.t. a designated reference: formation time in the left panel and momentum scheme in the right panel. }
\label{fig:nsplits}
\end{figure}

Next, we examine the number of emissions generated along the quark branch, shown in figure~\ref{fig:nsplits}. These distributions provide a proxy for the average length of the parton cascade, thus connecting to our stopping condition criteria. They also offer a first insight into potential differences between shower setups. While the overall shape of the distributions remains similar across configurations, we observe that showers ordered according to invariant mass tend to exhibit slightly more splittings. This is attributed to the energy independence of the corresponding stopping condition $s_{\min} = 4\Lambda^2$, in contrast to the energy-suppressed thresholds present in formation time and angular ordering. Consequently, the available phase-space for splittings depletes more slowly in the mass-ordered case, allowing the shower to evolve further before reaching the non-perturbative cutoff. Differences arising from the choice of kinematic scheme remain of similar magnitude, with the largest discrepancies confined to the high-multiplicity tail of the distribution that is less frequent in typical parton shower events.

\begin{figure}[ht]
\centering
\includegraphics[width=1.00\textwidth]{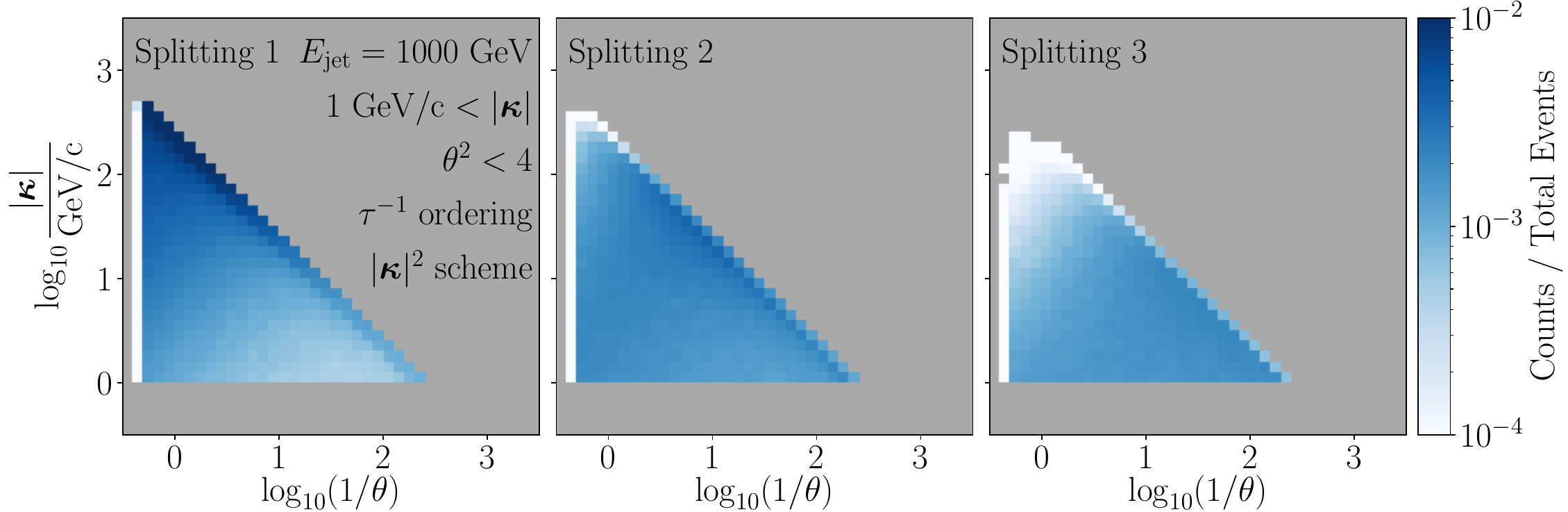}
\includegraphics[width=1.00\textwidth]{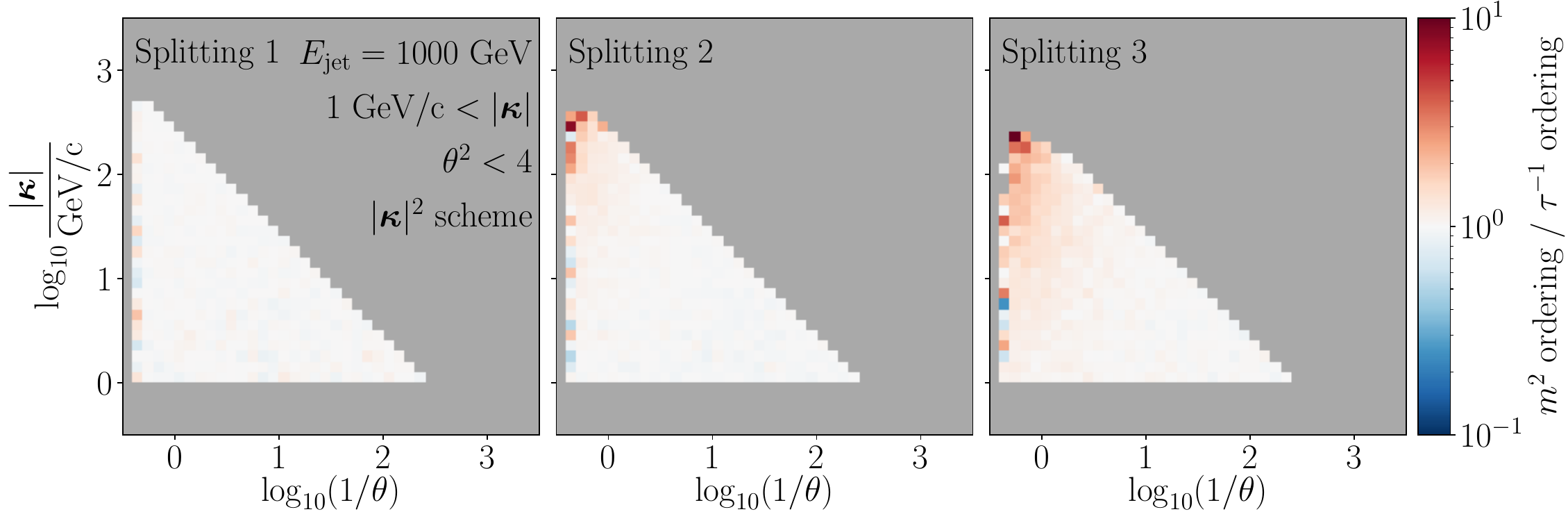}
\includegraphics[width=1.00\textwidth]{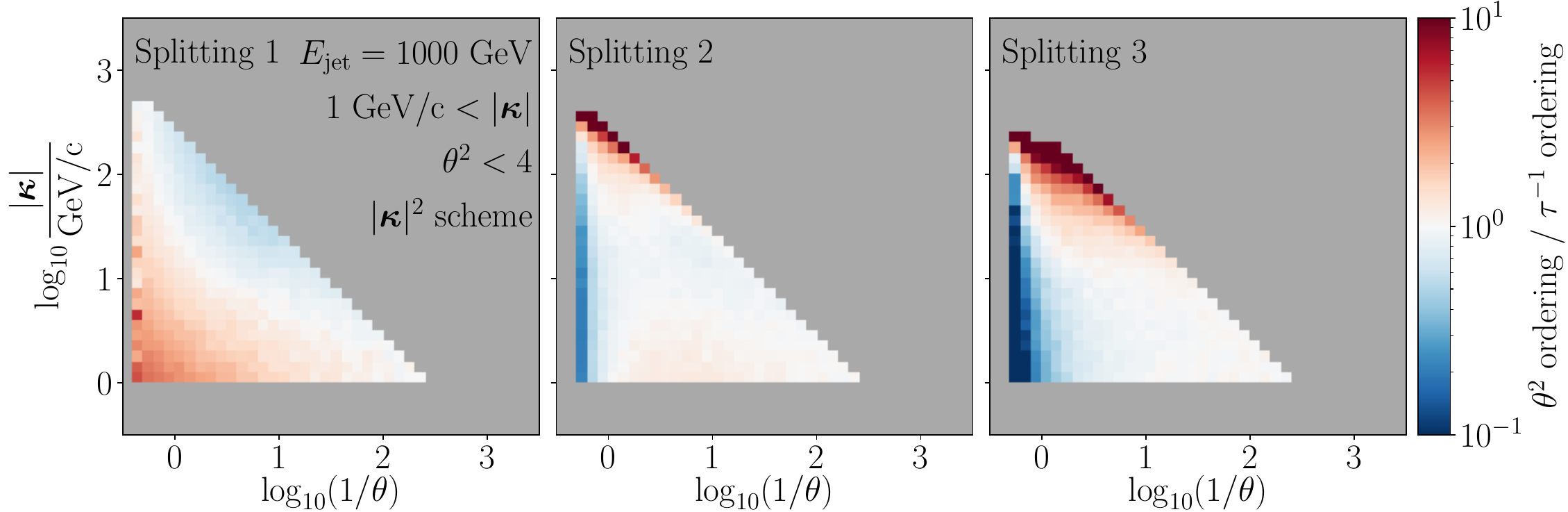}
\caption{Lund plane distributions in the $(\angleVAR, \ptrel)$ configuration for the first three $q\to qg$ splittings of cascades generated according to the momentum scheme. \textbf{Top:} Densities for $\invtf$ ordered cascades. Bins below the minimum are shown in white and empty bins shown in grey. All densities are normalised against the total number of events. \textbf{Middle:} Ratio between $\mass^2$ and $\invtf$ ordered cascades. \textbf{Bottom:} Ratio between $\angleVAR^2$ and $\invtf$ ordered cascades.}
\label{fig:lundplane-algoratios}
\end{figure}

To provide a more differential view of these effects, we turn to the Lund plane distributions associated with the first three emissions in the cascade. In figure~\ref{fig:lundplane-algoratios}, we show the results obtained within the momentum scheme for the three ordering prescriptions. Namely, the first row refers to the results obtained with the $\invtf$, while the middle and bottom rows refer to the ratios of the $\mass^2$ and $\angleVAR^2$-ordered showers to the $\invtf$-ordered case respectively. We observe that all three configurations populate broadly the same phase-space, bounded by the perturbative cutoff $|\ptrel| > \Lambda$, the initial angular condition\footnote{This bound arises from requiring the formation time \(\tform \sim 1/(z(1-z) E \theta^2)\) of each emission to be larger than the inverse of the parton energy. Since \(z(1-z) \leq 1/4\), the angular constraint \(\theta^2 < 4\) ensures that emissions occur well after the hard scattering event and are causally ordered.} $\theta^2 < 4$, and the kinematic limit $z(1-z) \leq 1/4$. However, subtle differences emerge in how this phase-space is filled: angular ordering tends to produce emissions closer to the soft and wide-angle region, while formation-time and mass ordering show a preference for more collinear and harder emissions. These differences, although formally subleading at DLL accuracy, may still translate into meaningful physical consequences when embedded in a medium. 

\begin{figure}[ht]
\centering
\includegraphics[width=.45\textwidth]{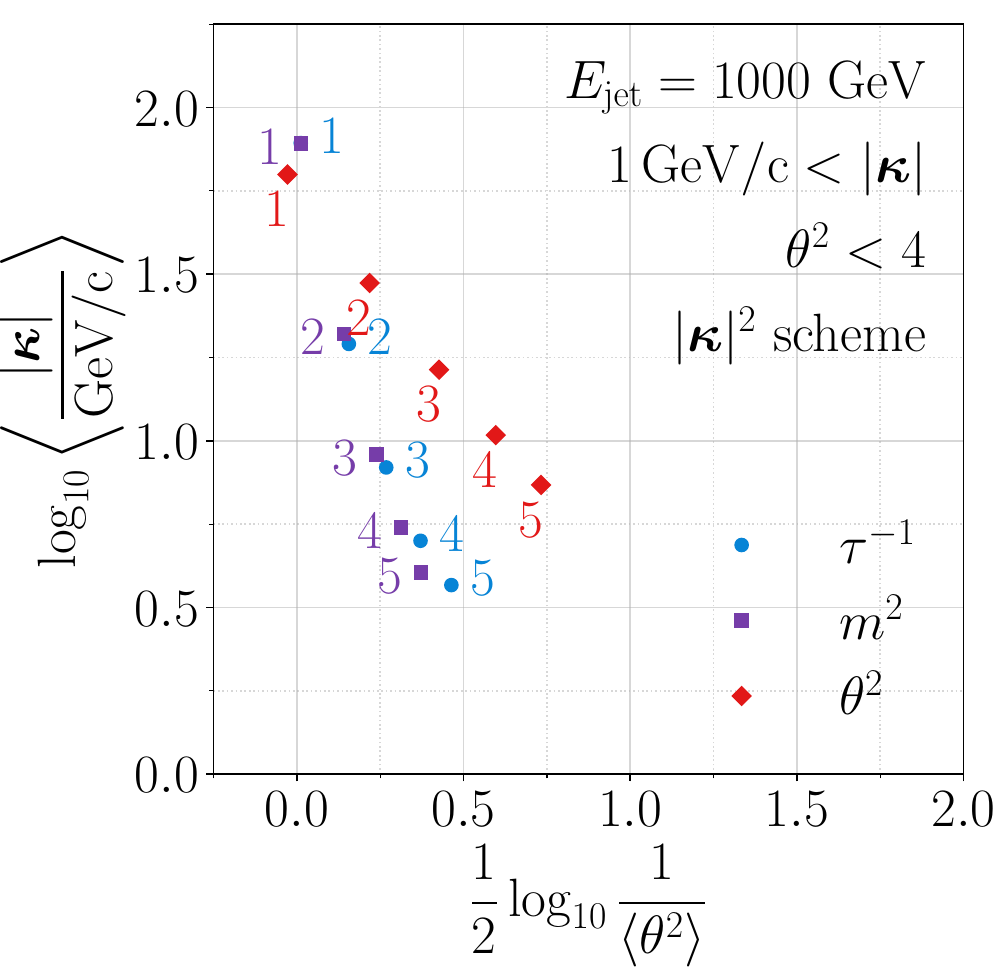}
\hspace{0.05\textwidth}
\includegraphics[width=.45\textwidth]{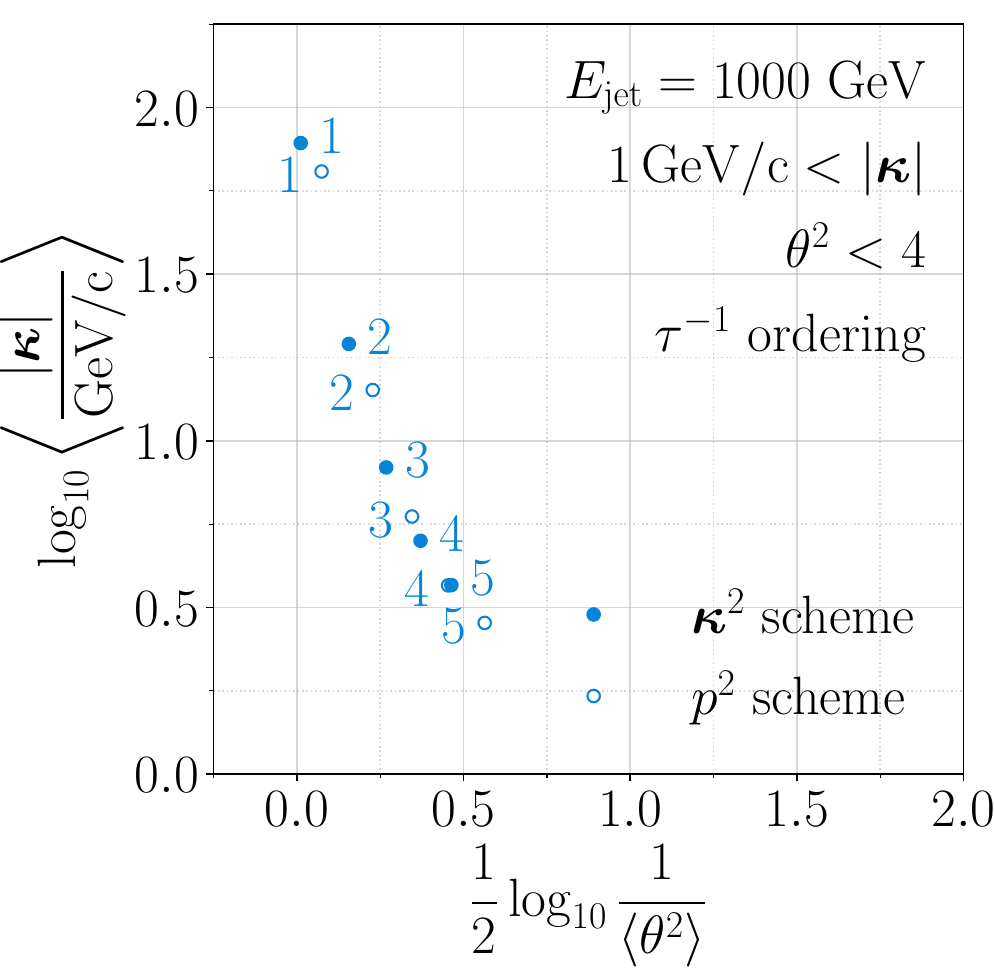}
\caption{Average Lund plane trajectories: \textbf{Left:} in the momentum kinematic scheme for $\invtf$ (blue circles), $\mass^2$ (purple squares), and $\angleVAR^2$ (red diamonds) ordering variables; \textbf{Right}: two reconstruction schemes, the momentum-scheme (solid blue circles) and mass-scheme (open blue circles) for the $\invtf$ ordering. The values were computed for the first five quark-initiated splittings, and are labelled in the corresponding colours.}
\label{fig:lund-trajectories}
\end{figure}

The potential impact of these differences becomes evident when we compare the phase-space trajectories of the showers across different ordering prescriptions (left) and kinematic schemes (right), as shown in figure~\ref{fig:lund-trajectories}. Despite the general consistency of their momentum structure, the average values from formation time and angular opening differ from one configuration to another, while the differences due to the choice of kinematic scheme are more modest. This reinforces the need to carefully account for these theoretical uncertainties when interfacing parton showers with jet quenching models, especially when formation time is used as a proxy for real time.

Overall, we demonstrate that while the choice of kinematic scheme has a relatively minor impact on the global features of the parton shower, the choice of ordering variable plays a more significant role in shaping its internal structure and, crucially, its space-time evolution. These findings will serve as a baseline for the discussion in the next section, where we explore how these uncertainties propagate to energy loss studies under the influence of a QCD medium.

\section{Energy Loss Effects}
\label{sec:energyloss}

Jets traversing a QCD medium are subject to substantial modifications due to interactions with the surrounding environment. These modifications are collectively referred to as jet quenching~\cite{Mehtar-Tani:2013pia,Blaizot:2015lma,Apolinario:2022vzg} and encompass a wide range of phenomena including energy loss, broadening, and decoherence. From a theoretical standpoint, a central aspect of jet quenching is the competition between the dynamical evolution of the parton shower and the finite extent of the medium. In particular, the ability of the medium to resolve individual parton splittings depends on the formation time of those splittings and on the medium resolution scale, often characterized by its length or coherence time. In this section, we aim to quantify how the uncertainties inherent to vacuum parton shower modelling, specifically those tied to the choice of ordering variable, propagate to predictions for in-medium energy loss. To this end, we embed our vacuum parton shower in a simplified static medium and analyse the resulting sensitivity of jet quenching observables to different shower configurations.

\subsection{Energy Loss Model}
\label{sec:quenching-model}

Our simplified quenching prescription is based on a constant density medium with fixed length $L$ and a formation-time criterion. Specifically, each emission is assigned a formation time $\tform$, and is considered to occur outside the medium if $\tform > L$. Conversely, emissions with $\tform < L$ are regarded as fully formed within the medium. The formation time $\tform$ for each emission is computed using the ordering variable and kinematic scheme described in section~\ref{sec:toymc-ordering}. In both the momentum and mass scheme, the formation time is given by
\begin{equation}
\tform = \frac{E z(1-z)}{|\ptrel|^2}\, ,
\end{equation}
where $E$ is the energy of the emitter, $z$ is the momentum fraction of the emitted gluon, and $|\ptrel|$ is the relative transverse momentum between the two daughter partons. This expression is consistent with uncertainty principle arguments and is widely used in phenomenological applications. In section~\ref{sec:energyloss-hi}, we compare the results obtained from both schemes. 

Once each emission is assigned a formation time, the quenching condition is implemented by comparing $\tform$ to the medium length $L$. In our model, this is interpreted as a direct mapping between formation time and real time, identifying $\tform$ as the time at which the emission is resolved. This identification is commonly used in Monte Carlo generators such as JEWEL~\cite{Zapp:2008gi,Zapp:2013vla}, MATTER~\cite{Majumder:2013re}, and the Hybrid model~\cite{Casalderrey-Solana:2014bpa}, and underlies their evaluation of medium properties at the time of splitting. A notable exception is JetMed~\cite{Caucal:2018dla,Caucal:2020zcz}, which employs angular ordering for vacuum emissions and light-cone time ordering for medium-induced splittings, reflecting distinct physical mechanisms.

We simplify the jet-medium interactions into a series of phase-space cuts. In particular, we adopt a minimal colour-coherence picture that includes a condition for the medium to resolve a parton splitting. This resolution depends on the relation between the formation time and the decoherence time $t_{\text{dec}}$, defined by
\begin{equation}
\hat{q} \tform > |\ptrel|^2 \quad \Longleftrightarrow \quad \tform > \left(\hat{q} \angleVAR^2\right)^{-1/3} \equiv t_{\text{dec}}(\angleVAR)\,,
\end{equation}
where $\hat{q}$ is the medium transport coefficient and $\angleVAR^2$ is the squared opening angle of the splitting. Both conditions imply that $\angleVAR^2 > \theta_\text{dec}^2 = 1/(\hat{q}L^3)$~\cite{Mehtar-Tani:2011vlz,Casalderrey-Solana:2012evi}, ensuring that the splitting is sufficiently wide-angled for the medium to resolve its colour structure. These two criteria -- a formation time smaller than the medium length and the decoherence phenomena -- are combined into a single quenching condition,
\begin{equation}
\mathcal{Q} = \Theta(t_{\text{dec}} < \tform < L)\,,
\label{eq:quench-prob}
\end{equation}
where $\Theta$ is the Heaviside function. This formulation ensures that only emissions occurring within the medium and within the decoherence timescale are \textit{quenched}. We consider two possible strategies for applying the quenching condition:
\paragraph{First Splitting:} the jet is discarded if the first sampled emission is inside the phase-space region defined by the quenching condition~\eqref{eq:quench-prob};
\paragraph{Full Shower:} the jet is discarded if any splitting along the quark branch satisfies the quenching condition.
\paragraph{}

This setup enables a clean comparison between shower configurations. In the next sections, we will investigate how differences in the ordering variable and kinematic scheme affect quenching magnitudes under this model.

\subsection{Uncertainties for Jet Energy Loss Studies in Heavy-ion Collisions}
\label{sec:energyloss-hi}

We now examine how the simplified quenching model described above leads to different amounts of lost jets depending on the underlying parton shower configuration. For this purpose, we introduce a single quantity: the fraction of \textit{quenched} events in the sample, defined as
\begin{equation}
    \frac{N_{\text{quenched}}}{N_{\text{vacuum}}} \times 100\,\% \,,
\label{eq:quench}
\end{equation}
where the numerator counts the number of events
discarded using the considerations in the previous subsection,
and the denominator is the total number of vacuum events generated. 

\begin{figure}[ht]
\centering
\includegraphics[width=.45\textwidth]{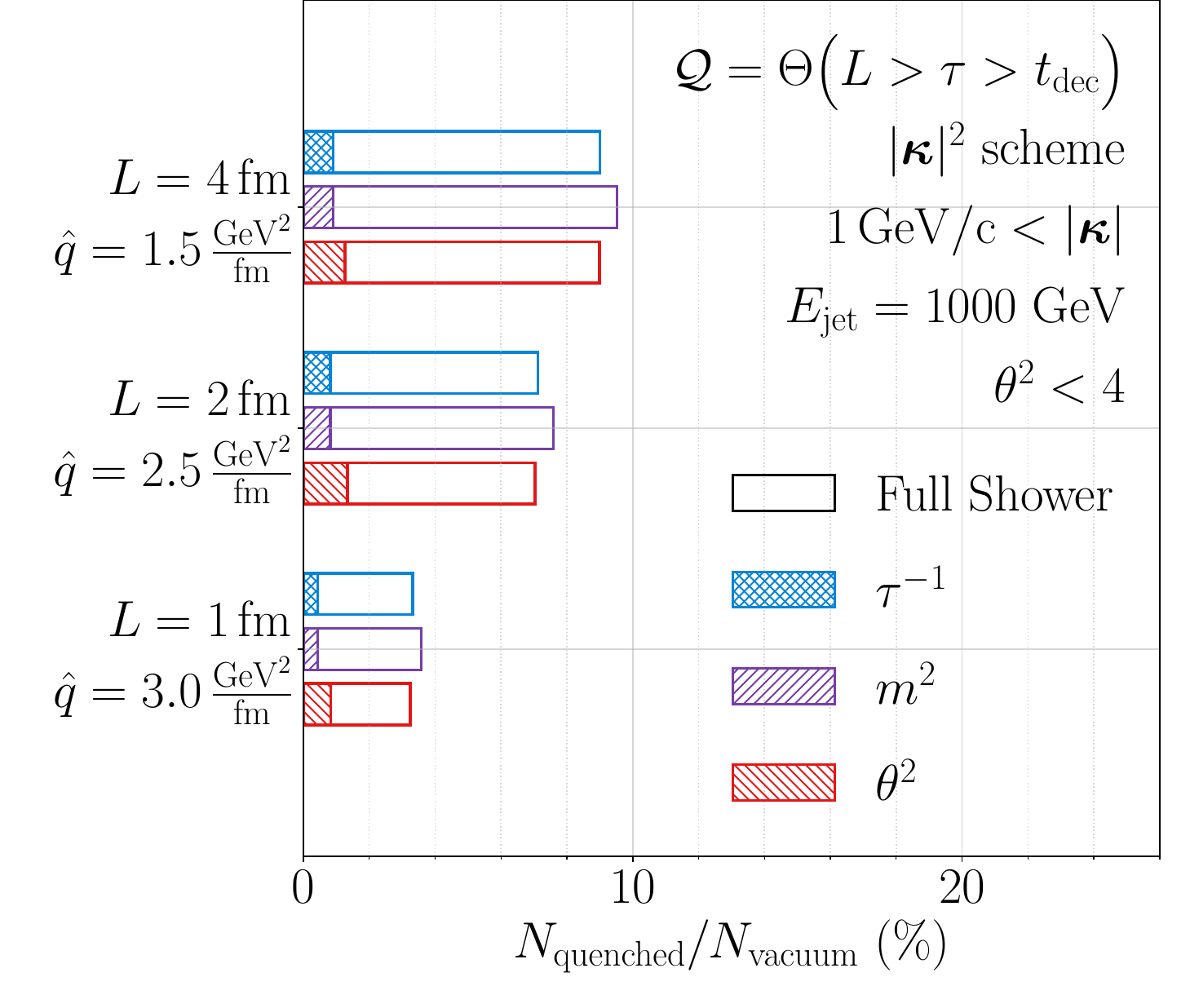}
\includegraphics[width=.45\textwidth]{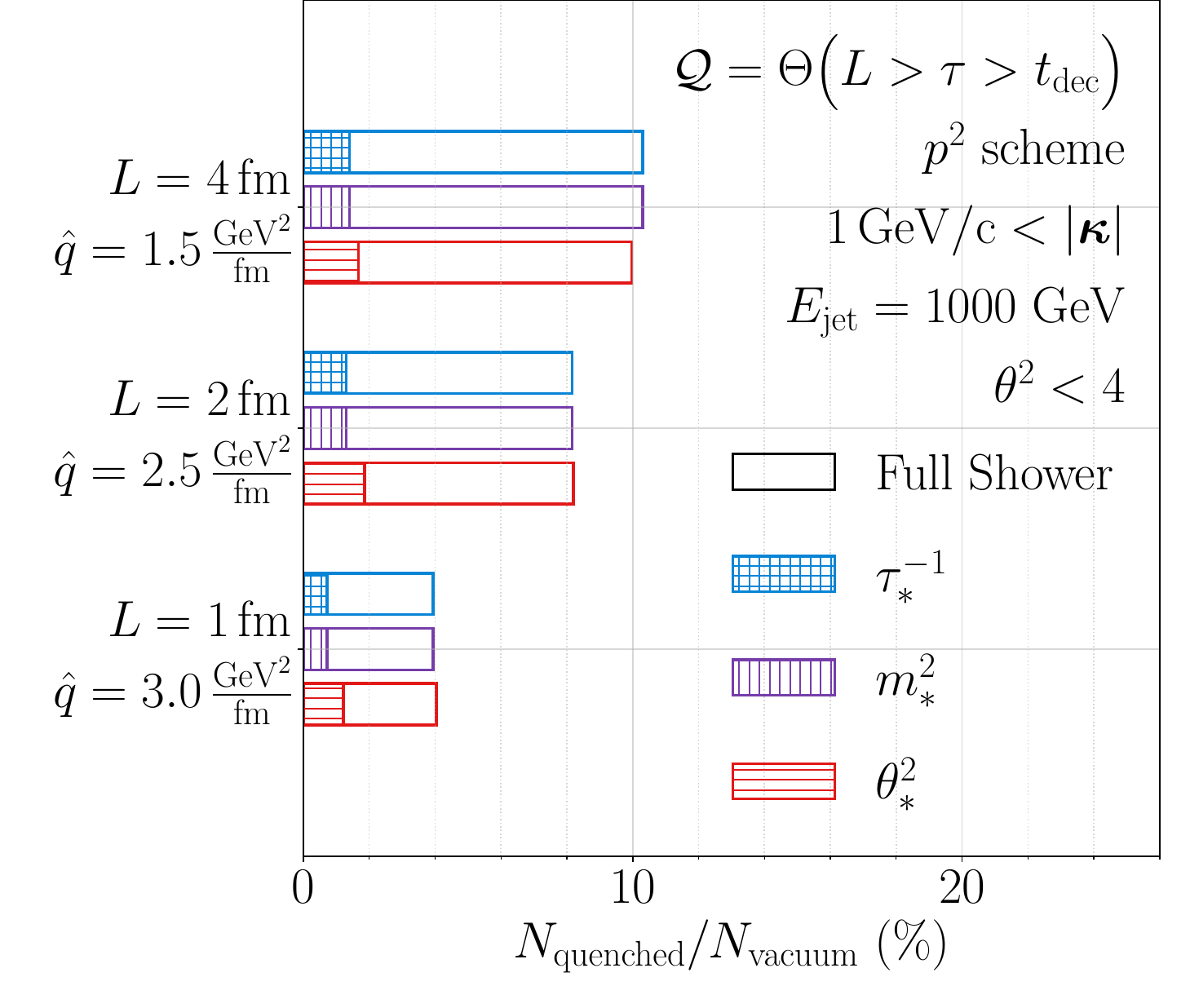}
\caption{Percentage of events obeying the quenching condition with different values of $(L, \hat{q})$ in eq.~\eqref{eq:quench-prob} in samples generated according to formation time (blue crossed), invariant mass (purple slanted), and angle (red counter-slanted) ordering prescriptions. Hatched rectangles correspond to the ``First Splitting'' mode and empty rectangles to the ``Full Shower''. Vacuum samples are built out of events generated in the momentum scheme (left panel), and mass scheme (right panel).}
\label{fig:quench-weights}
\end{figure}

We compute this quantity for all three ordering prescriptions ($\invtf$, $\mass^2$, and $\angleVAR^2$), as well as their proxies in the case of the mass scheme,  under both the “First Splitting” and “Full Shower” quenching modes introduced previously. As shown in figure~\ref{fig:quench-weights}, differences in the amount of quenched jets \eqref{eq:quench} emerge even in this idealized model. The left panel illustrates the effect for the momentum reconstruction scheme, while the right panel refers to the mass kinematic scheme where $\tau^{-1}$ and $m^2$ ordering coincide. %
These differences are most pronounced for short-lived media (small $L$), and in the “First Splitting” mode, where angular ordering results in slightly larger quenching probabilities than the other cases. This behaviour correlates with the tendency of angular-ordered showers to populate the soft and wide-angle region of the Lund plane, which is more likely to fall within the quenched region of phase-space.

\subsubsection{Structure of the Quenched phase-space}
\label{sec:quench-phase-space}

To understand the interplay between vacuum shower dynamics and medium-induced suppression, it is useful to examine how the quenching condition \eqref{eq:quench-prob} selects specific regions of the Lund plane. Since this condition depends explicitly on both the formation time and the opening angle of each splitting, it defines a characteristic region of phase-space (bounded by the medium length and coherence effects) where the medium can significantly alter the parton cascade.

We begin by recalling the quenching condition, eq.~\eqref{eq:quench-prob}, where the decoherence time $t_{\rm dec}$ depends on the emission angle through $t_{\rm dec}(\angleVAR) = (\hat{q}\angleVAR^2)^{-1/3}$, and $L$ is the total in-medium path length. This imposes a double-sided constraint in the Lund plane:

\begin{itemize}
\item \textbf{Lower bound:} $\tform > t_{\rm dec}(\angleVAR)$, defining the condition for vacuum-like emissions to be resolved by the medium.
\item \textbf{Upper bound:} $\tform < L$, setting the usual space-time boundary for in-medium emissions.
\end{itemize}

\begin{figure}[ht]
\centering
\includegraphics[width=0.45\textwidth]{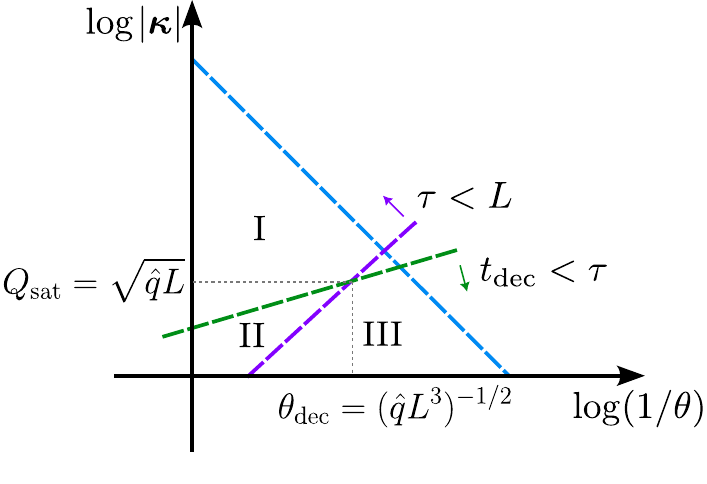}
\includegraphics[width=0.45\textwidth]{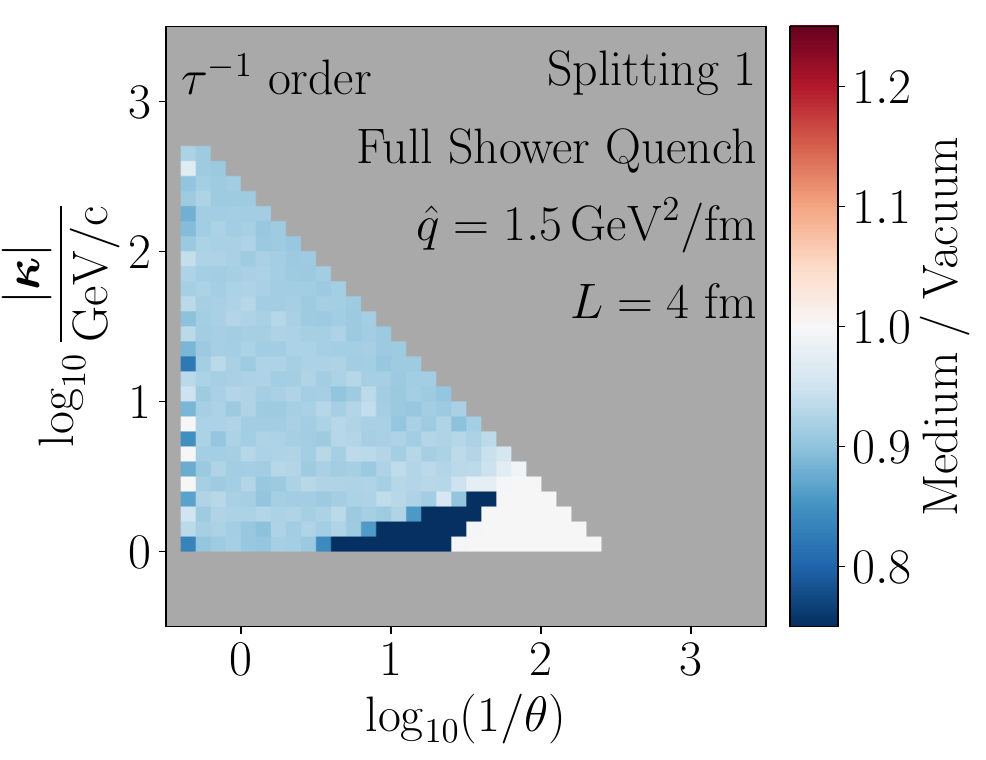}
\caption{
\textbf{(Left)} Different phase-space regions relevant for the pseudo-quenching models, including in-medium vacuum-like emissions (region I), broadening dominated splittings (region II), and splittings outside of the medium (region III). \textbf{(Right)}
Ratio between `quenched' and `vacuum' Lund densities of the first splitting along the quark branch for $\tform^{-1}$ ordered cascades. The quenching parameters are $L= \SI{4}{fm}$ and $\hat{q} = \SI{1.5}{GeV/fm^2}$ when applying the model while following the ``Full Shower" strategy.}
\label{fig:quench-phase-space}
\end{figure}

Together, these conditions imply a minimum angle for resolved splittings, defined as $\theta^2_\text{dec} = 1/(\hat{q}L^3)$.
The left panel of figure~\ref{fig:quench-phase-space} illustrates the quenching-relevant regions in the $(\log(1/\theta), \log |\ptrel|)$ plane. The medium-length constraint $\tform < L$ is shown as a purple dashed line, while the decoherence time condition $t_{\rm dec} = (\hat{q}\angleVAR^2)^{-1/3}$ is represented by a green dashed curve. The region above the purple line, consisting of regions I and II, contains splittings with formation times shorter than the medium length and thus potentially affected by the medium. Region III, below the purple line, corresponds to splittings that occur entirely outside the medium and are unaffected. Within regions I and II, the subset of splittings that also satisfy the decoherence condition, that is, those resolved by the medium, are indicated as Region II. This region defines the effective phase-space where medium-induced modifications, such as energy loss, can occur within our simplified quenching model.

For illustrative purposes, the right panel of figure~\ref{fig:quench-phase-space} shows the ratio between `quenched' and `vacuum' Lund densities of the first quark splitting in $\invtf$ ordered cascades in the momentum scheme, when the ``Full Shower" strategy is adopted. We note that region II is completely eliminated, since vacuum-like splittings with such kinematics are forbidden, while events whose first splitting lie in region I may be depleted since subsequent splittings may belong to region II.

This figure reveals  the size and location of the quenched region sensitive to both the medium properties ($\hat{q}$ and $L$) and the angular structure of the vacuum cascade. In particular, we note that  shorter media or more collimated showers result in fewer splittings meeting the quenching condition, while broader cascades, such as those produced by angular ordering, are more likely to probe the decohered phase-space.

Moreover, this framework highlights the importance of early emissions: if the first splitting lies outside the quenched region, subsequent emissions have less chance of being quenched because the available phase space shrinks quickly. This motivates the distinction between ``First Splitting" and ``Full Shower" prescriptions when evaluating $N_{\rm quenched}/N_{\rm vacuum}$, as discussed in the previous section.

In the next section, we explore how formation time inversions and their treatment impact this quenching region and its associated suppression patterns.

\subsubsection{Mapping between formation time and in-medium path-length}
\label{sec:time-violations}

A key feature of this parton shower formulation framework is the possible mapping between formation time and real time, which enables a clear identification of emissions occurring inside or outside the medium. However, this mapping can be invalidated in some parton shower configurations due to the occurrence of \textit{formation time inversions}: situations where a later splitting in the sequence has a shorter formation time than an earlier one. Such inversions violate causality in the real-time picture and complicate the interpretation of the reconstructed $\tau$ as a true time ordering variable for other ordering variables.

In the simplified quenching model described above, the identification of the parton formation time with a real-time variable is key to determine whether a given splitting occurs inside the medium or not. However, in Monte Carlo implementations of parton showers, especially those involving mass- or angle-based orderings, this mapping is not always monotonic. This means that later emissions in the branching sequence may have smaller formation times than earlier ones, leading to \emph{time-ordering violations}. These violations are artefacts of the probabilistic generation of the splitting sequence based on the ordering variable $s$, which is not necessarily equivalent to the formation time. For instance, in a shower ordered by invariant mass $m^2$, a later branching with small $m^2$ but high energy may have a shorter formation time than its predecessor. As a result, the association between the shower depth and propagation time through the medium becomes ambiguous, especially in the ``Full Shower" quenching scenario.

\begin{figure}[ht]
\centering
\includegraphics[width=.475\textwidth]{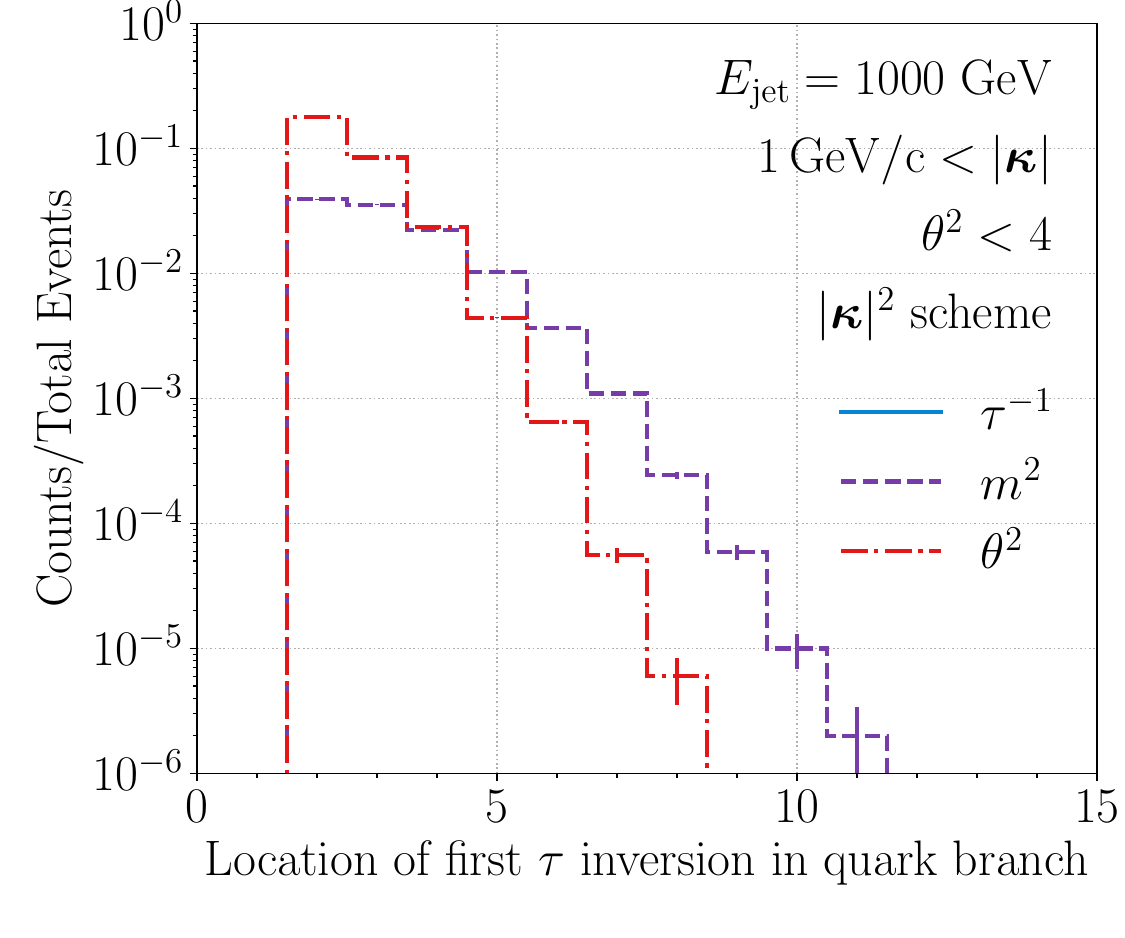}
\caption{Probability of the first time inversion occurring at a given depth $n$ in the parton cascade for different ordering prescriptions. Results are shown for the momentum kinematic scheme.}
\label{fig:time-inversion-freq}
\end{figure}

To better quantify these effects, we present in figure~\ref{fig:time-inversion-freq} the probability that the first time ordering violations occurs at any given point in the cascade, shown for each ordering prescription. We observe that the frequency of inversions decreases with shower depth and depends strongly on the ordering variable: while formation-time ordered showers trivially produce no inversions by construction, mass- and angle-ordered showers can exhibit inversion rates between $10\%$ and $30\%$ of the generated events. This highlights the need to properly handle such configurations when using formation time as a proxy for in-medium propagation.

To address this issue, we implement two mitigation strategies:

\begin{enumerate}
\item \textbf{Post-hoc removal of time-inverted events:} In this approach, we discard events from the vacuum sample if any inversion in the formation time sequence is detected along the quark branch. While simple, this introduces a selection bias. The removal of inverted-time events artificially skews the phase-space (see appendix~\ref{app:time-inversions}), particularly enhancing late-time, wide-angle emissions in angular-ordered showers.
\item \textbf{Veto-resampling procedure:} A more controlled alternative is to resample events such that all splittings are generated under a strict formation time constraint: $\tau_{n} > \tau_{n-1}$ for each subsequent branching. If a sampled emission violates this ordering, it is vetoed and resampled with updated bounds. This algorithm preserves the statistical integrity of the full vacuum sample while maintaining causal consistency in time, although differences in the Lund planes still persist (see appendix~\ref{app:time-inversions}).
\end{enumerate}

\begin{figure}[ht]
\centering
\includegraphics[width=0.800\textwidth]{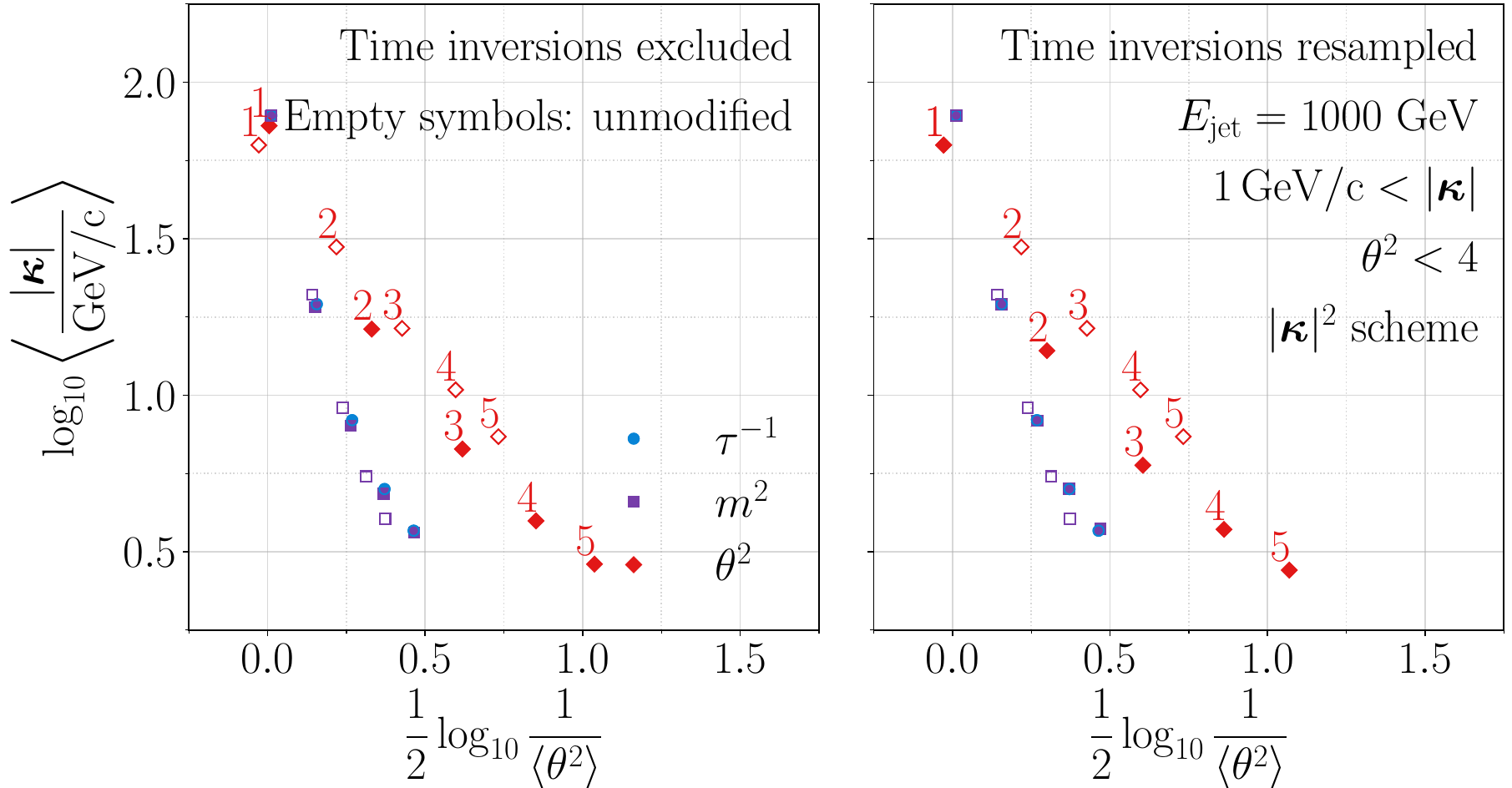}
\caption{Average Lund plane trajectories in the momentum kinematic scheme when time-inverted emissions are addressed using: \textbf{Left:} Post-hoc removal; \textbf{Right:} Veto-resampling procedure. The ordering variables $\invtf$, $\mass^2$ and $\angleVAR^2$ are respectively represented by blue circles, purple squares and red diamonds. In both panels empty markers correspond to the unmodified sample (i.e., with time inversions).
}
\label{fig:lund-trajectories-time-alt1}
\end{figure}

Now we examine the effect of such strategies on the splitting distributions.
The corresponding Lund plane trajectories for the momentum kinematic scheme are shown in the different panels of figure~\ref{fig:lund-trajectories-time-alt1}, where the trajectories of the unmodified samples (cf. figure~\ref{fig:lund-trajectories}) have been overlayed as empty markers. Both procedures result in a shift to later formation times, with similar effect on the Lund distribution average values. Nonetheless, the actual distributions are affected in different ways by the two removal strategies (see appendix \ref{app:time-inversions}), and change from the unmodified samples (therefore, from the DLL limit) are sizeable. 

\begin{figure}[ht]
\centering
\includegraphics[width=1.00\textwidth]{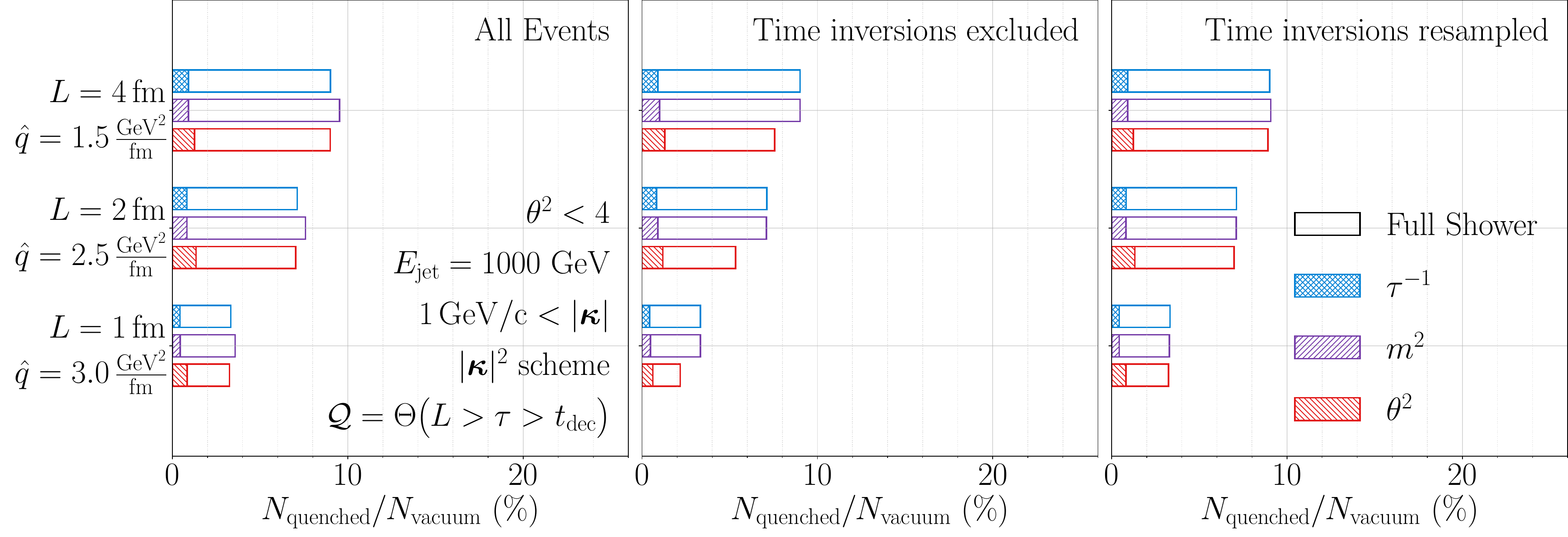}
\caption{Percentage of events obeying the quenching condition~\eqref{eq:quench-prob} with different values of $(L, \hat{q})$  for samples generated according to formation time (blue crossed), invariant mass (purple slanted), and angle (red counter-slanted) ordering prescriptions. Hatched rectangles correspond to the ``First Splitting" mode and empty rectangles to the ``Full Shower". Vacuum samples consisting of all events (left panel), only events with no inversions (centre panel), and inversion-resampled events (right panel).}
\label{fig:quench-weights-time}
\end{figure}

These methods define two alternative “vacuum” samples: one where only causally ordered events are retained, and one where the time ordering is built in from the outset. We apply the quenching condition to these samples and show the resulting $N_\text{quenched}/N_\text{vacuum}$ distributions in the middle and right panels of figure~\ref{fig:quench-weights-time}. The left panel corresponds to the full vacuum sample (same as figure~\ref{fig:quench-weights} for reference). The middle panel shows the result after removing inverted-time events, while the right panel corresponds to the veto-resampled sample. The discrepancies observed between these panels, particularly in the ``Full Shower" mode, demonstrate that formation time violations can significantly affect quenching probabilities, and that the resampling procedure is effective in mitigating these artifacts.

We emphasize that while the ``First Splitting" mode remains largely unaffected by time inversions since the first emission is rarely affected by reordering, their impact is substantial in multi-emission observables.  In effect, in the ``Full Shower” mode  differences arise, especially in the post-hoc exclusion case, and angular-ordered showers are more sensitive to inversions due to their broader angular phase-space. Therefore, in any in-medium application of parton showers, ensuring a consistent space-time picture is essential. The time veto procedure removes this sensitivity by construction, providing a more robust baseline (although some caveats remain, see appendix~\ref{app:angle-inversions}). In the following, unless otherwise specified, we adopt the veto-resampled samples for all medium studies, although for the  reasons mentioned, a more firmly grounded strategy would be desirable.

An analogous analysis can be performed for angular inversions, where the emission sequence violates monotonicity in the opening angle. The results, shown in appendix~\ref{app:angle-inversions}, indicate similar discrepancies, reinforcing the need to treat such ordering violations carefully when using an ordering variable as proxy for real-time evolution.

In summary, even within our highly idealized model, the choice of ordering variable and the treatment of time-ordering violations cause  variations in quenching predictions. These differences highlight the importance of incorporating consistent space-time constraints into parton shower algorithms, particularly when interfacing with jet quenching models in heavy-ion collisions.

\subsubsection{Alternative Energy Loss Models}
\label{sec:eloss-alternatives}

While our primary quenching model uses a sharp condition on formation time and decoherence (eq.~\eqref{eq:quench-prob}), realistic jet–medium interactions are expected to be more gradual and model-dependent. In this subsection, we explore alternative energy loss models by systematically varying the quenching condition, and testing sensitivity to jet energy and hadronisation scale. This analysis allows us to assess the robustness of our findings regarding the impact of the parton shower configuration.

We begin by considering two alternative quenching conditions in addition to the one introduced in section~\ref{sec:quenching-model}:
\begin{equation}
\begin{aligned}
\mathcal{Q} &= \Theta(\tform < L) \cdot \exp \left\{- \tform^3 / t_{\rm dec}^3 \right\} \,, 
\\
\mathcal{Q} &= \Theta(\tform < L) \cdot \Theta(\hat{q}L > |\ptrel|^2) \,.
\end{aligned}
\label{eq:extra-quenching-conditions}
\end{equation}
The first introduces a smooth suppression factor reflecting a decoherence rate-inspired quenching probability~\cite{Mehtar-Tani:2011vlz}, while the second imposes a sharp transverse momentum cutoff below the saturation scale of the medium $Q^2_{\rm sat} = \hat{q}L$. Both conditions differ from the default model in their treatment of soft and wide-angle emissions, with the latter condition more aggressively removing splittings in the low-$|\ptrel|$ region of the Lund plane.

To mimic the presence of a jet cone, we introduce a cut on the splitting angle, considering only events where $\angleVAR < R_{\rm max}$ for all $q \to qg$ splittings in the quark branch. This restricts the analysis to the collinear region of the shower, helping to disentangle these emissions from (less well controlled) wide-angle contributions. 

\begin{figure}[ht]
\centering
\includegraphics[width=1.00\textwidth]{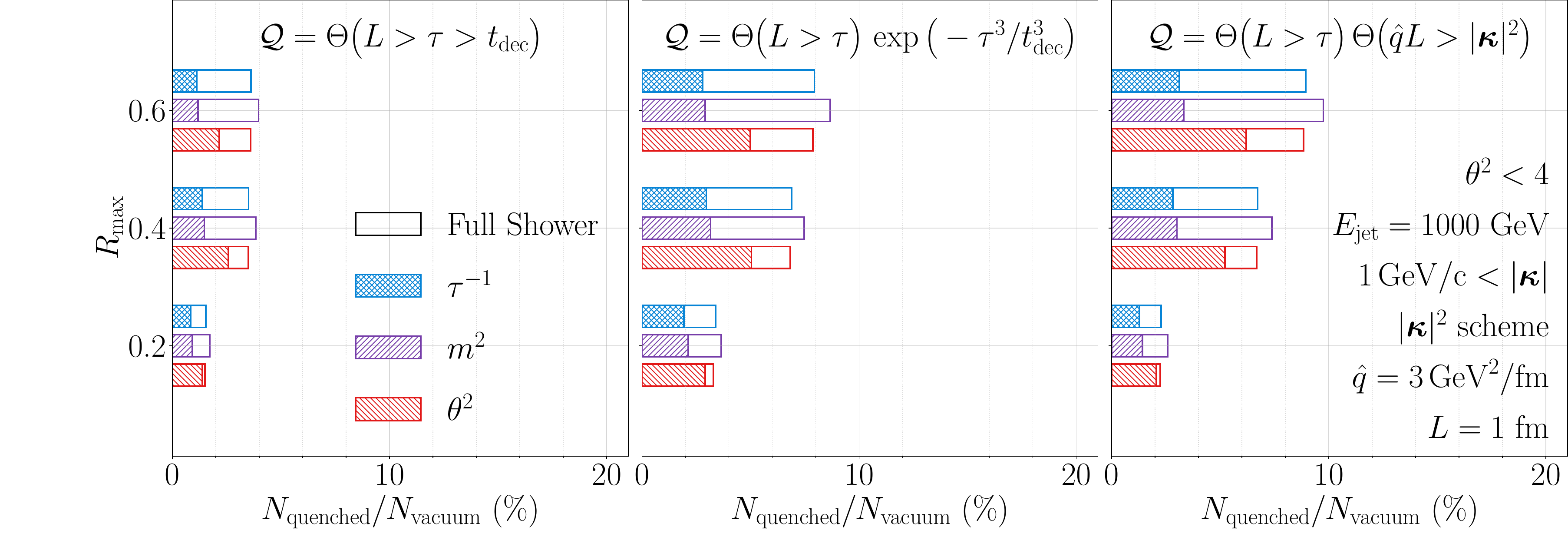}
\includegraphics[width=1.00\textwidth]{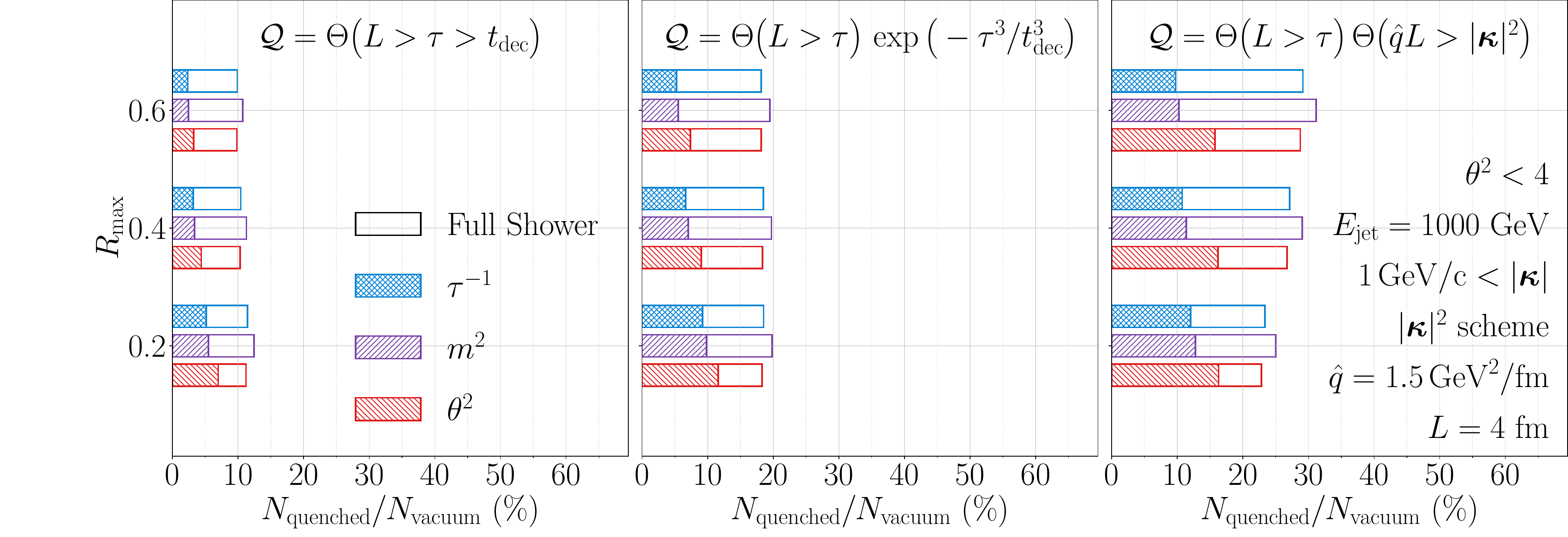}
\caption{Percentage of events obeying the quenching conditions in eq.~\eqref{eq:extra-quenching-conditions} in samples generated according to formation time (blue), invariant mass (purple), and angle (red) ordering prescriptions. The Hatched rectangles correspond to the ``First Splitting'' mode, while the empty rectangles to the ``Full Shower''. All samples were restricted such that quark-branch splittings had $\angleVAR < R_{\rm max}$ before quenching. The top panel corresponds to ($L=\SI{1}{fm}$, $\hat{q}=\SI{3}{GeV^2/fm}$), and the bottom panel to ($L=\SI{4}{fm}$, $\hat{q}=\SI{1.5}{GeV^2/fm}$). The quenching model alternatives are indicated in the plots, and each plot shows results from increasing values of $R_{\rm max}$ from bottom to top. }
\label{fig:quench-weights-modelcomp}
\end{figure}

Figure~\ref{fig:quench-weights-modelcomp} shows the quenched fractions for all three quenching models, comparing ``First Splitting'' and ``Full Shower’’ prescriptions for three values of $R_{\rm max}$ and two medium scenarios. The top panel represents a short-lived, dense medium ($L = 1$ fm, $\hat{q} = 3$ GeV$^2$/fm), while the bottom panel corresponds to a long-lived, dilute one ($L = 4$ fm, $\hat{q} = 1.5$ GeV$^2$/fm). The dependence on the ordering variable persists across all configurations, reinforcing its role as a key source of uncertainty. Interestingly, the observed trend with $R_{\rm max}$ depends on the medium profile: for short-lived media, the ``Full Shower'' quenched fraction increases with $R_{\rm max}$, while for long-lived media, the ``First Splitting’’ quenched fraction decreases. This interplay arises because wider jets are more likely to include early, soft emissions (which dominate quenching in compact media), but also exhibit shorter decoherence times (which dominate quenching in extended media). The second model in eq.~\eqref{eq:extra-quenching-conditions}, being independent of $t_{\rm dec}$, displays a distinct behaviour. In appendix~\ref{app:extra-quenching-results} (figure~\ref{fig:quenching-ratios-schemecomp}) we repeat this exercise while varying the kinematic scheme, and note how its impact on the quenching fraction is significantly smaller than that of the ordering prescription, especially in the collinear region.

\begin{figure}[ht]
\centering
\includegraphics[width=1.00\textwidth]{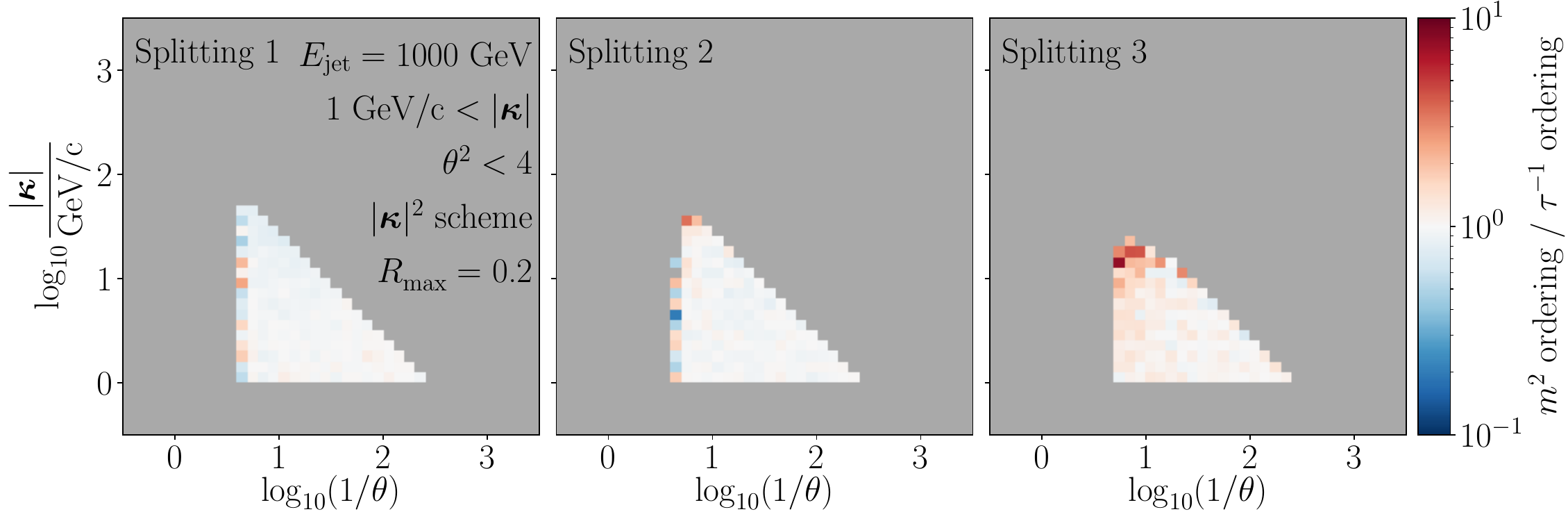}
\includegraphics[width=1.00\textwidth]{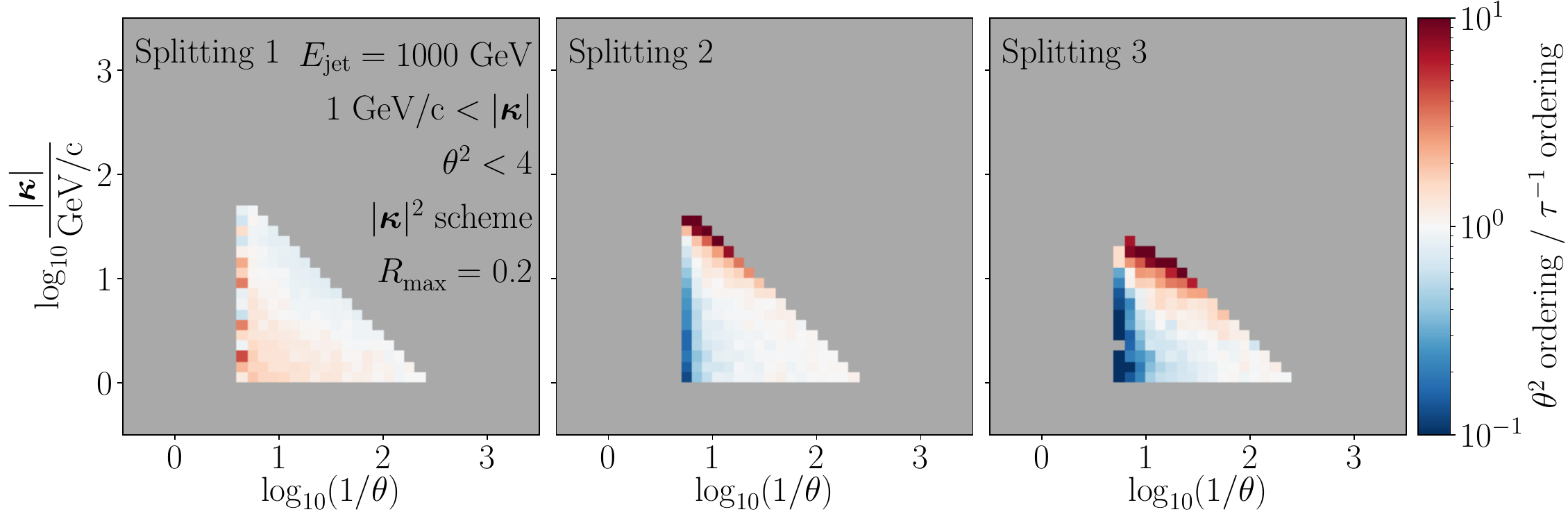}
\caption{Lund plane distributions in the $(\angleVAR, \ptrel)$ configuration for the first three $q\to qg$ splittings of cascades generated according to the momentum scheme. The samples restricted to only those events where all $\angleVAR < R_{\rm max} = 0.2$. \textbf{Top:} Ratio between $\mass^2$ and $\invtf$-ordered cascades. \textbf{Bottom:} Ratio between $\angleVAR^2$ and $\invtf$-ordered cascades.}
\label{fig:lundplane-algoratios-Rmax02}
\end{figure}

To further clarify why differences between algorithms persist even when angular cuts are imposed, figure~\ref{fig:lundplane-algoratios-Rmax02} shows the Lund plane ratios for $\mass^2$- and $\angleVAR^2$-ordered showers relative to $\invtf$-ordered ones, after applying $R_{\rm max} = 0.2$. While the wide-angle region is excluded, algorithmic differences re-emerge near the angular cutoff, suggesting that the showers are effectively rescaled versions of one another. Thus, restricting the angular phase-space does not eliminate the imprint of the ordering choice.

We also tested an alternative angular restriction, where only the splittings exceeding a fixed angle $r_g$ are discarded (while the event is retained), which slightly modifies the parton shower history. This approach yielded results consistent with the hard $R_{\rm max}$ cut, but led to marginally smaller quenched fractions.

Finally, we explored the influence of jet energy and hadronisation scale, presented in appendix~\ref{app:extra-quenching-results}. Overall, we find that lower hadronisation scales enhance the total number of splittings and increase the observable differences between ordering prescriptions, especially in the ``Full Shower'' mode. On the other hand, reducing the jet energy tends to decrease the absolute number of emissions, diminishing the differences between the ``First Splitting’’ and ``Full Shower’’ quenching modes. Importantly, the observed trends with the ordering variable remain robust under these variations.

\section{Conclusions}
\label{sec:conclusions}

In this work, we have presented a systematic investigation of how technical aspects of parton shower implementations specifically the choice of ordering variable and kinematic kinematic scheme, affect the structure of jet evolution and its interplay with a QCD medium. This framework is built upon a simplified Monte Carlo model at double logarithmic accuracy, which allows for transparent comparisons between different configurations while preserving the essential features of QCD radiation.

In vacuum, we showed that while final-state momentum distributions remain largely consistent across different ordering prescriptions, sizeable differences emerge in the space-time development of the shower. In particular, formation time trajectories differ depending on whether the shower is ordered in inverse formation time, virtuality, or angle. These variations, although formally subleading, become highly relevant when interfacing the parton shower with a medium.

To explore the consequences for energy loss studies, we implemented a minimal model for jet quenching based on colour decoherence and formation-time constraints. Within this model, we quantified the fraction of quenched events under different quenching prescriptions and showed that the choice of ordering variable significantly impacts the extent and location of the quenched phase-space. These differences persist even when the analysis is restricted to the collinear region of the Lund plane. We also addressed the issue of formation time inversions, which arise in showers not ordered in $\tform$ and undermine the interpretation of formation time as a physical time variable. We showed that such inversions can skew quenching observables, particularly in multi-branch emissions. Two mitigation strategies, post-hoc event removal and a veto-based resampling procedure, were proposed and evaluated, with the latter offering a robust way to restore time ordering while maintaining statistical consistency.

Finally, we examined how the results are affected by modifying the quenching model itself. Alternative quenching prescriptions, such as continuous suppression or transverse momentum-based constraints, were tested alongside with variations in jet energy and hadronisation scale. Across all scenarios, the dependence on the parton shower configuration remained significant, reinforcing our conclusion that the treatment of space-time structure in parton showers is a major source of uncertainty in jet quenching studies.

This work highlights the importance of incorporating consistent space-time dynamics in future parton shower developments. In particular, the formation time should be treated not only as a useful diagnostic tool but also as a guiding principle in constructing realistic jet quenching frameworks. Our findings provide a quantitative baseline for assessing such improvements and emphasize the need for careful theoretical control over the microscopic evolution of jets in heavy-ion collisions.

\acknowledgments

This work is supported by European Research Council project ERC-2018-ADG-835105 YoctoLHC, by OE - Portugal, Fundação para a Ciência e a Tecnologia (FCT), I.P., under ERC-PT A-Projects ``Unveiling", by Xunta de Galicia (CIGUS Network of Research Centres), by European Union ERDF, and by the Spanish Research State Agency under projects PID2020-119632GBI00 and PID2023-152762NB-I00. This work is part of the project CEX2023-001318-M financed by MCIN/AEI/\-10.13039/\-501100011033. LA, AC, and CA acknowledge support by FCT under contracts 2021.03209.\-CEECIND, PRT/BD/154190/2022 and 2023.07883.CEECIND respectively. The work of CA was partially supported by the U.S. Department of Energy, Office of Science, Office of Nuclear Physics under grant Contract Number DE-SC0011090 and by FCT, I.P., project 2024.06117.CERN.

\appendix

\section{Time Inversions}
\label{app:time-inversions}

\begin{figure}[ht]
\centering
\includegraphics[width=\textwidth]{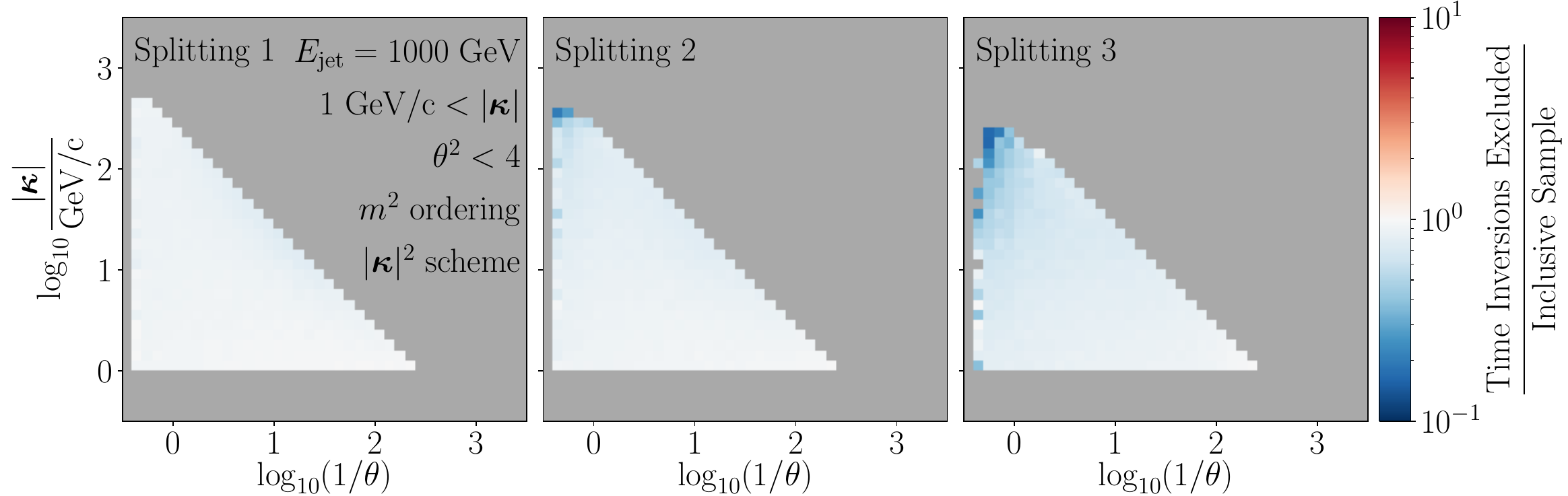}
\includegraphics[width=\textwidth]{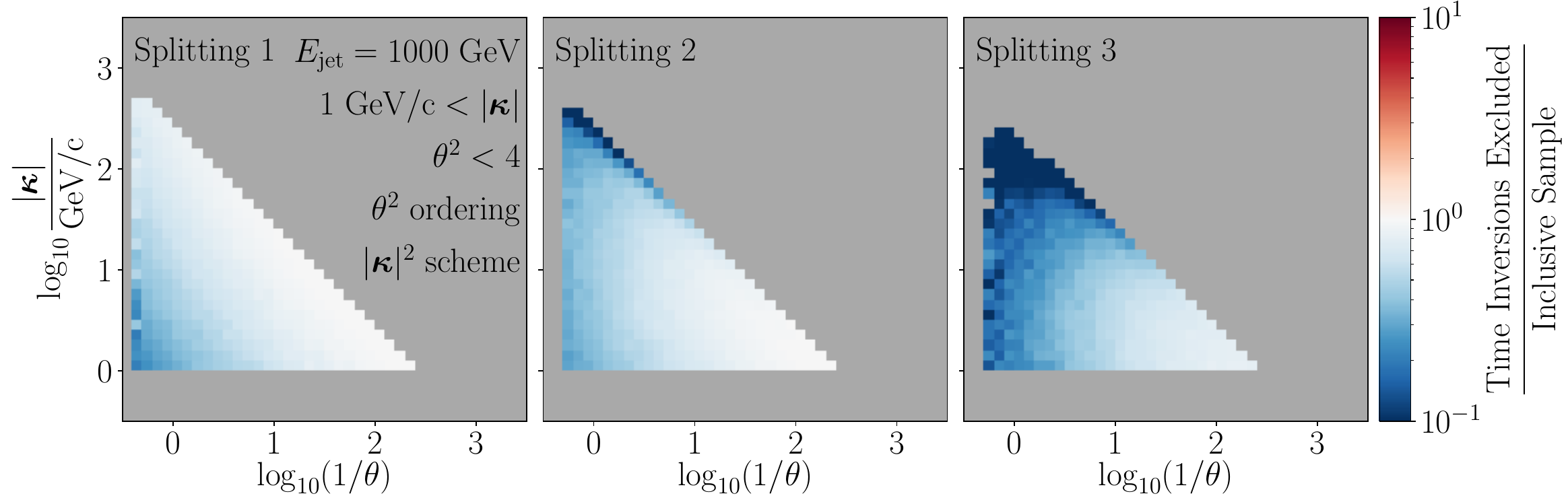}
\caption{Lund plane distributions of the first three splittings for the sample with a \textit{post-hoc} exclusion of formation time (see subsection~\ref{sec:time-violations}) inversions divided by the corresponding Lund distributions for the inclusive samples, for cascades ordered in $\mass^2$ \textbf{(top)}, and $\angleVAR^2$ \textbf{(bottom)} in the momentum kinematic scheme.}
\label{fig:timeinvcomp-noinversions}
\end{figure}

The significant fraction of events with at least one formation time inversion in the quark branch raises the question of how these inversions affect the shower substructure. To explore the potential interface between the shower and the evolving medium, we adopt the two strategies for eliminating formation time inversions discussed in the main text. First, we simply exclude all events with at least one time inversion in their quark branch. The resulting Lund distributions are presented in figure~\ref{fig:timeinvcomp-noinversions} as ratios to the unmodified (or inclusive) samples. 

In these ratios, we observe minimal changes for the $\mass^2$-ordered cascades (top panel), with most of the modifications concentrated at wide angles and large $|\ptrel|$ values for the second and third splittings. This region corresponds to early values of $\tform$, which are more likely to exhibit inversions. For $\angleVAR^2$-ordered cascades (bottom panel), the effect is significantly more pronounced, with the first splitting distribution being strongly suppressed in the low-$z$ region, corresponding to events with a large remaining phase-space for the second splitting, which leads to a higher probability of time inversions. Additionally, the second and third splittings are suppressed for early formation times. In both cases, the \textit{post-hoc} exclusion of formation time inversions significantly alters the Lund distributions. These modifications are similar in magnitude to the differences between ordering prescriptions.

\begin{figure}[ht]
\centering
\includegraphics[width=\textwidth]{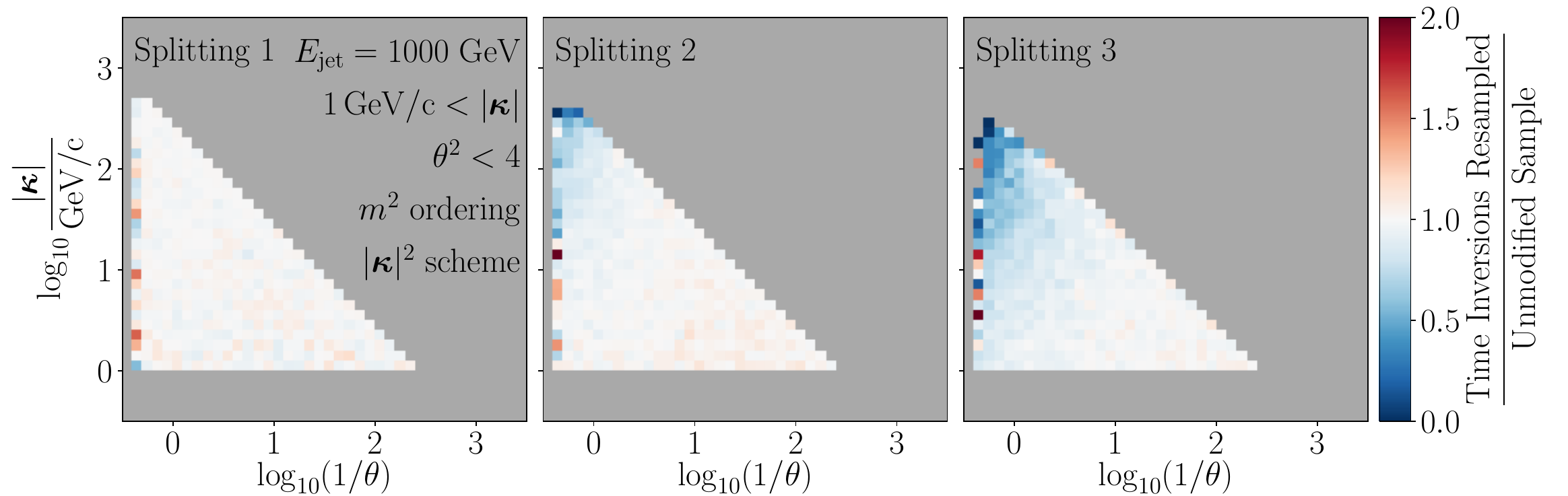}
\includegraphics[width=\textwidth]{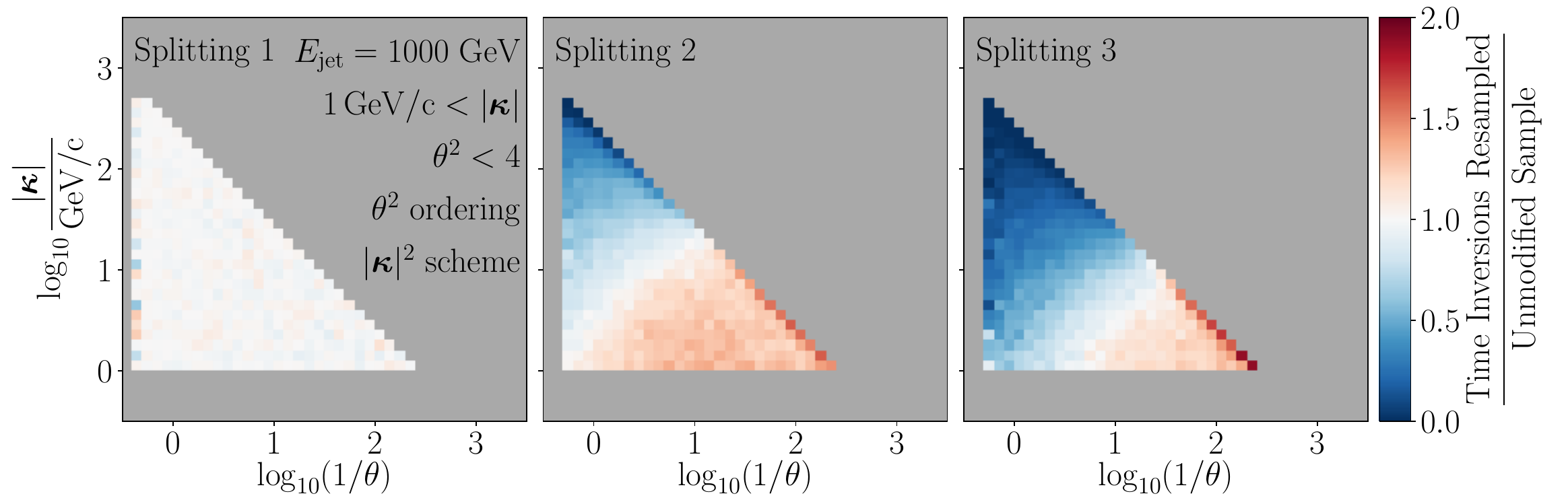}
\caption{Ratio between Lund plane distributions of the sample with resampled formation time inversions and the samples with unrestricted evolution, for cascades ordered in $\mass^2$ \textbf{(top)} and $\angleVAR^2$ \textbf{(bottom)} in the momentum kinematic scheme. The first three quark-branch splittings were considered.}
\label{fig:timeinvcomp-veto}
\end{figure}

The second approach to eliminating formation time inversions consists of implementing a veto and resampling procedure during the shower generation. In this method, the splitting scale ($\mass^2$ or $\angleVAR^2$) and splitting fraction ($z$) are sampled as described in subsection~\ref{sec:toymc-algo}, with the additional step of evaluating the formation time $\tform$ and comparing it to that of the previous splitting. If the trial splitting results in a time inversion, it is rejected, and the generation continues with a lower scale until a suitable splitting is found. The resulting ``Time Inversions Resampled'' Lund distributions, shown in figure~\ref{fig:timeinvcomp-veto}, are presented as a ratio to the unmodified (inclusive) samples. 

In this approach, the first splitting distributions remain unaltered, since the no-inversion condition does not impose any restrictions on the first emission. However, the second and third splittings are noticeably modified. For $\mass^2$-ordered cascades the modification is concentrated in the wide-angle, large-$|\ptrel|$ corner of phase-space, with fewer modifications in the collinear region than the \textit{post-hoc} removal case. For $\angleVAR$-ordered showers, we observe a sharp division between depleted and enhanced regions of the Lund plane, with their boundary given by a line of constant formation time,  which increases with each splitting. 
While this resampling-based implementation controls the modifications to the shower substructure better than the \textit{post-hoc} removal of time inversions, the Lund distributions are still significantly altered, further increasing the inherent  uncertainty to the double logarithmic approximation.

We note that the results here refer solely to the momentum reconstruction scheme, with inversions being absent in the mass scheme (see the discussion in subsection~\ref{sec:toymc-algo}). Nonetheless, the latter is just one of several possible choices inherent to a parton shower, making the present study a relevant example of the non-trivial effects that can arise when removing time-inversions. For a further examination of the difference between the resampling procedure and post-hoc exclusions, as well as the possibility of eliminating angular ordering violations, see appendix~\ref{app:angle-inversions}.

\section{Angle Inversions}
\label{app:angle-inversions}

\begin{figure}[h!]
\centering
\includegraphics[width=.475\textwidth]{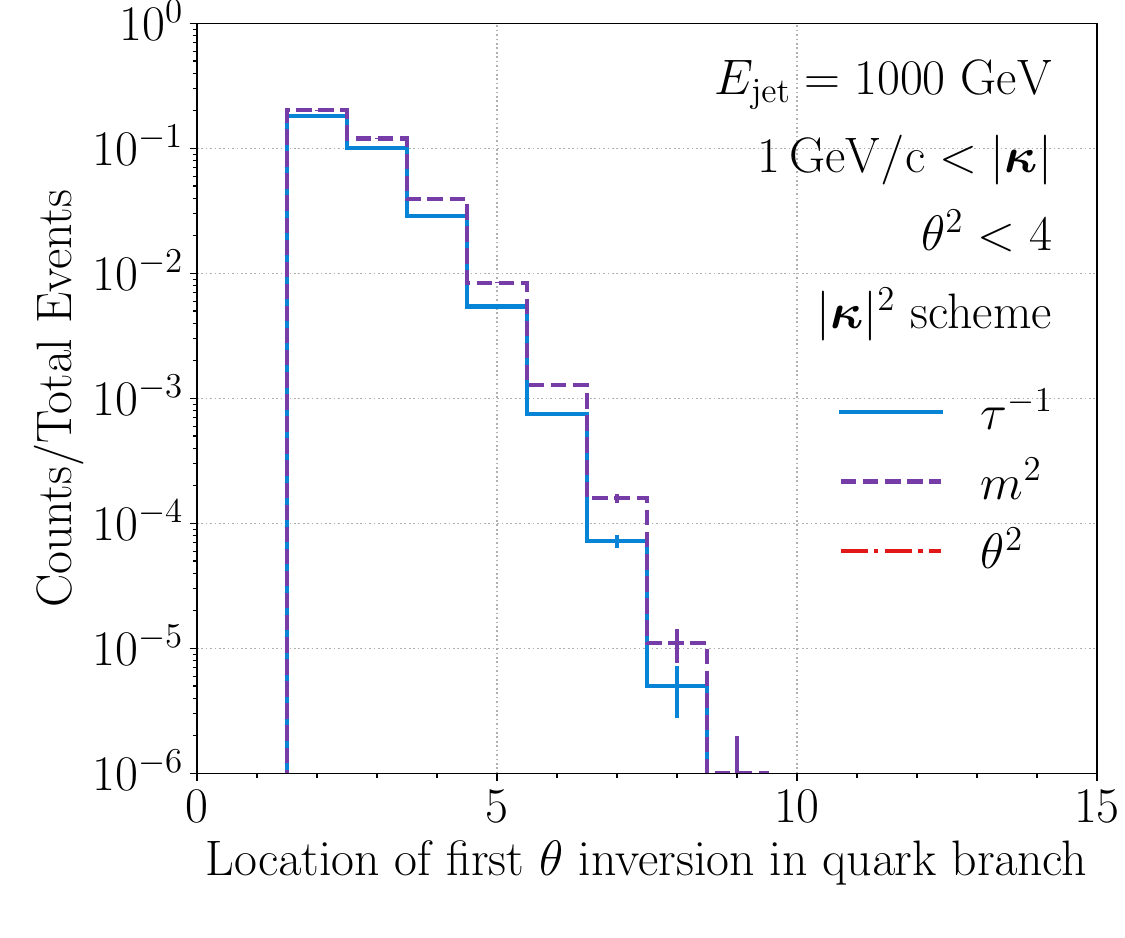}
\caption{Location of the first inversion in opening angle along the quark branch of parton cascades generated according to the three ordering prescriptions in the momentum scheme: $\invtf$ (blue solid line), $\mass^2$ (purple dashed line), and $\angleVAR^2$ (red dot-dashed line). Histograms expressed as fractions of the total number of events ($10^6$).
}
\label{fig:angle-inversions-algo-comparison}
\end{figure}

In this appendix we turn our attention to the effect of angular ordering violations and their removal, employing methods analogous to those used to address time ordering violations. Figure~\ref{fig:angle-inversions-algo-comparison} shows the probability for the first angular ordering violation to occur at any given point in the cascade. Much like for time ordering violations (cf. figure~\ref{fig:time-inversion-freq}), most angular inversions happen early in the quark branch, and $30\%$ of events show at least one angular inversion.

\begin{figure}[h!]
\centering
\includegraphics[width=1.00\textwidth]{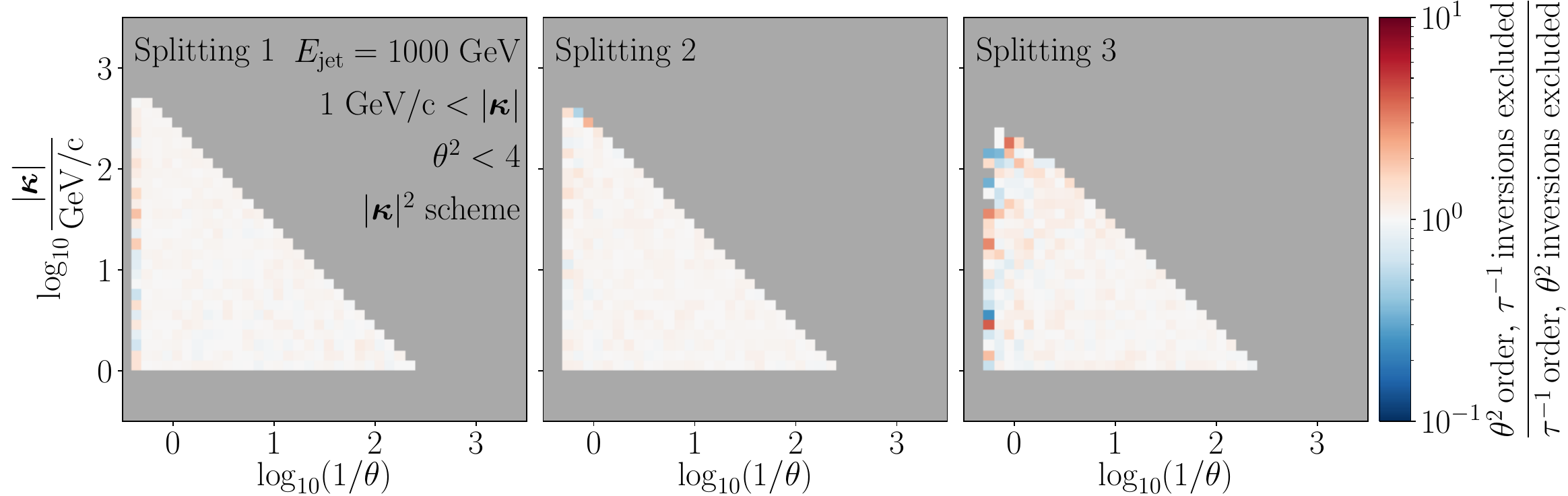}
\includegraphics[width=1.00\textwidth]{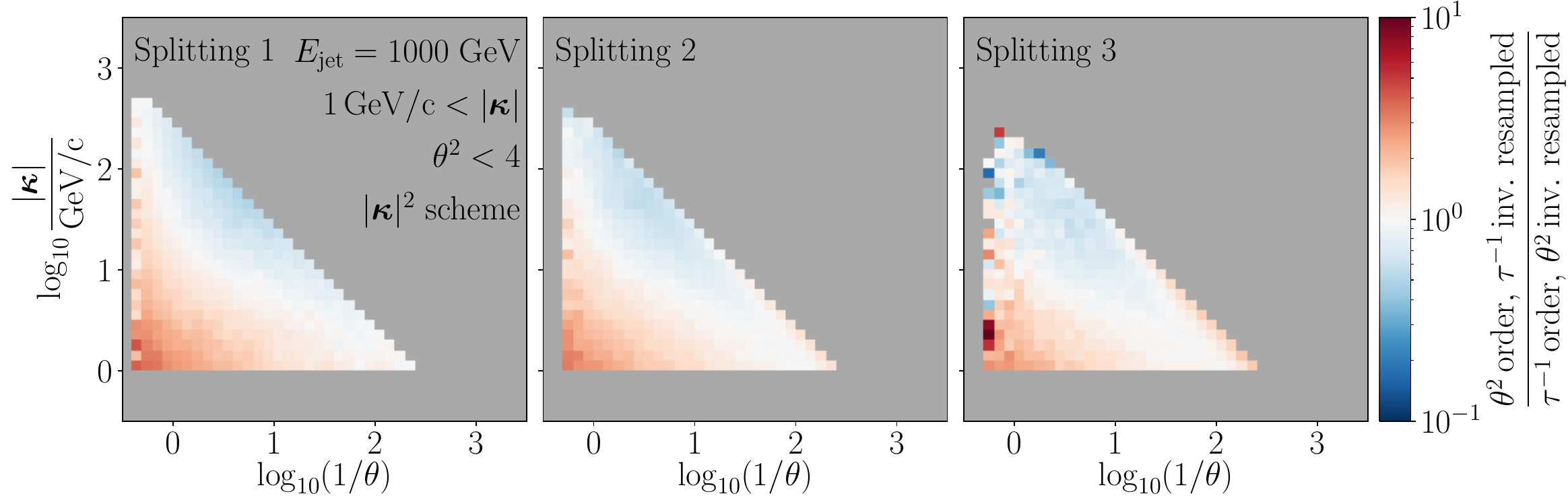}
\caption{Ratios of Lund plane densities for the three first quark-initiated splittings of cascades generated according to the ``$|\ptrel|^2$ scheme''. \textbf{Top:} Ratio between the $\invtf$-ordered sample with angular inversions excluded \textit{post-hoc} and the $\angleVAR^2$-ordered sample with time inversions excluded  \textit{post-hoc}. %
\textbf{Bottom:} Ratio between the $\invtf$-ordered sample where angular inversions are resampled and the $\angleVAR^2$-ordered sample where time inversions are resampled.}
\label{fig:veto-exclusion-comparison}
\end{figure}

The relative impact of angular versus time ordering violations can be assessed by comparing the effect of time inversion removal in an angular-ordered sample with that of removing angular inversions in a time-ordered sample. This is shown in figure~\ref{fig:veto-exclusion-comparison}. Specifically, the top panel shows  ratios between  two different samples: the formation time-ordered sample in which all events with at least one $\angleVAR^2$-ordering violation along the quark  are excluded, divided by the angular-ordered sample in which all events with at least one formation-time ordering violation along the quark branch are excluded. This ratio is compatible with unity across the entire phase-space for the first three splittings, with some statistical fluctuations observed in the case of the third splitting in the region of large angles and small formation times, where events are likely to be excluded from the numerator or denominator, respectively. We also note that this ratio tends to be slightly above unity, consistent with the slightly higher frequency of angular inversion in the $\invtf$-ordered sample ($\sim$ 37\%) compared to formation-time inversions in the $\angleVAR^2$-ordered sample ($\sim$ 29\%) (see figure~\ref{fig:time-inversion-freq}).

When repeating this exercise  using a veto procedure instead of a \textit{post-hoc} exclusion, that is by treating either time or angle inversions as unresolved splittings, one obtains the Lund ratios in the bottom panel of figure~\ref{fig:veto-exclusion-comparison}, which exhibit rather strong modifications, especially towards softer $z$ distributions. This analysis reveals that while the resampling procedure preserves the first splitting distributions with respect to the unmodified sample, the post-hoc exclusion procedure better respects the strong ordering in different variables. As such, \textit{there is no clear advantage to either method for preventing inversions in any given shower variable}.

\begin{figure}[h!]
\centering
\includegraphics[width=1.00\textwidth]{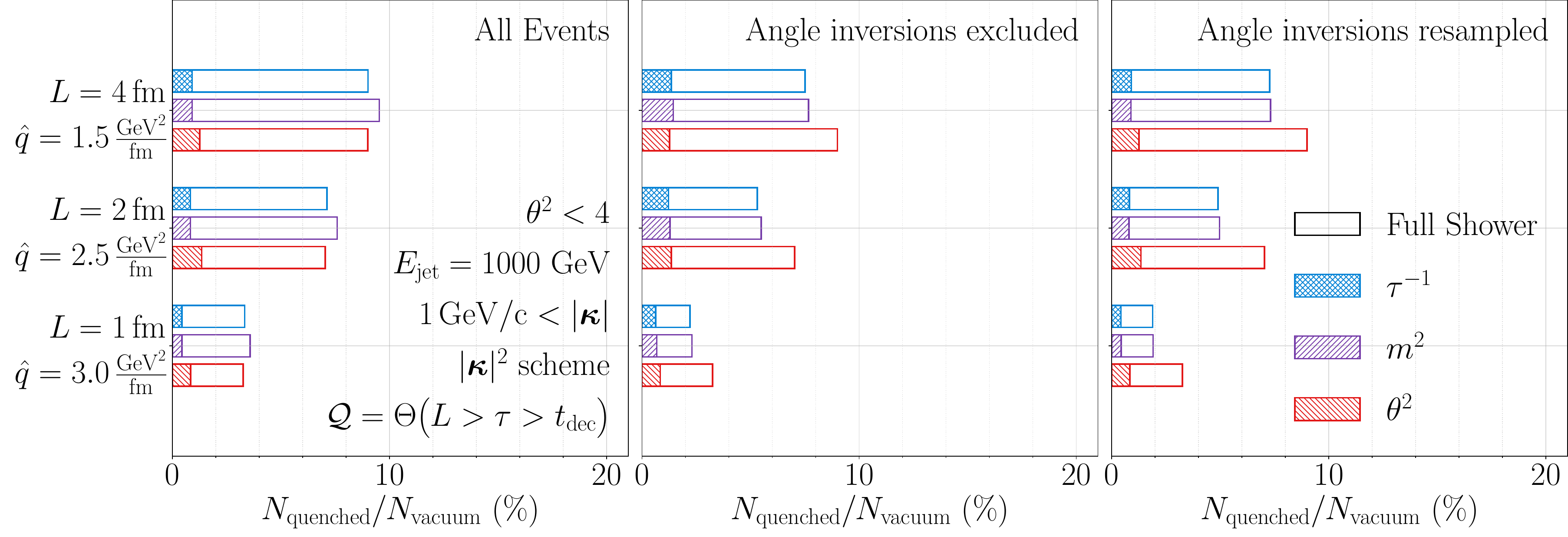}
\caption{Percentage of events satisfying the quenching condition in eq.~\eqref{eq:quench-prob} for different values of $L$ and $\hat{q}$ and for samples generated according to formation-time (blue), invariant-mass (purple), and angular (red) ordering prescriptions. Results on vacuum samples consisting of all events (left panel), events with  angular inversions excluded \textit{post-hoc} (centre panel), and  events with resampled angular inversions (right panel). Hatched rectangles correspond to the ``First Splitting'' mode, while empty rectangles to the ``Full Shower''.}
\label{fig:quench-weights-anglevetocomp}
\end{figure}

It is also worth considering the role of angular inversions in our pseudo-quenching model. To this end, we repeat the calculations that led to figure~\ref{fig:quench-weights}, but now preventing inversions in angle rather than in formation time. The results are shown in figure~\ref{fig:quench-weights-anglevetocomp}. We note that the source of differences between algorithms is not solely the presence of angular ordering, although it does contribute to the sensitivity to jet colour decoherence. When angular inversions along the quark branch are excluded \textit{post-hoc} (middle panel), significant differences between the different orderings are observed in the   ``Full Shower'' mode  (empty rectangles), as parton cascades are biased towards configurations where later splittings are more collinear. When these inversions are prevented by resampling (right panel), a similar behaviour emerges, although in this case the differences are already present in the ``First Splitting'' mode.
Overall, we see how a simple model for jet decoherence is sensitive to the choice of  ordering prescription and to how angular ordering is enforced.

\section{Further results in jet quenching}
\label{app:extra-quenching-results}

\begin{figure}[h!]
\centering
\includegraphics[width=1.00\textwidth]{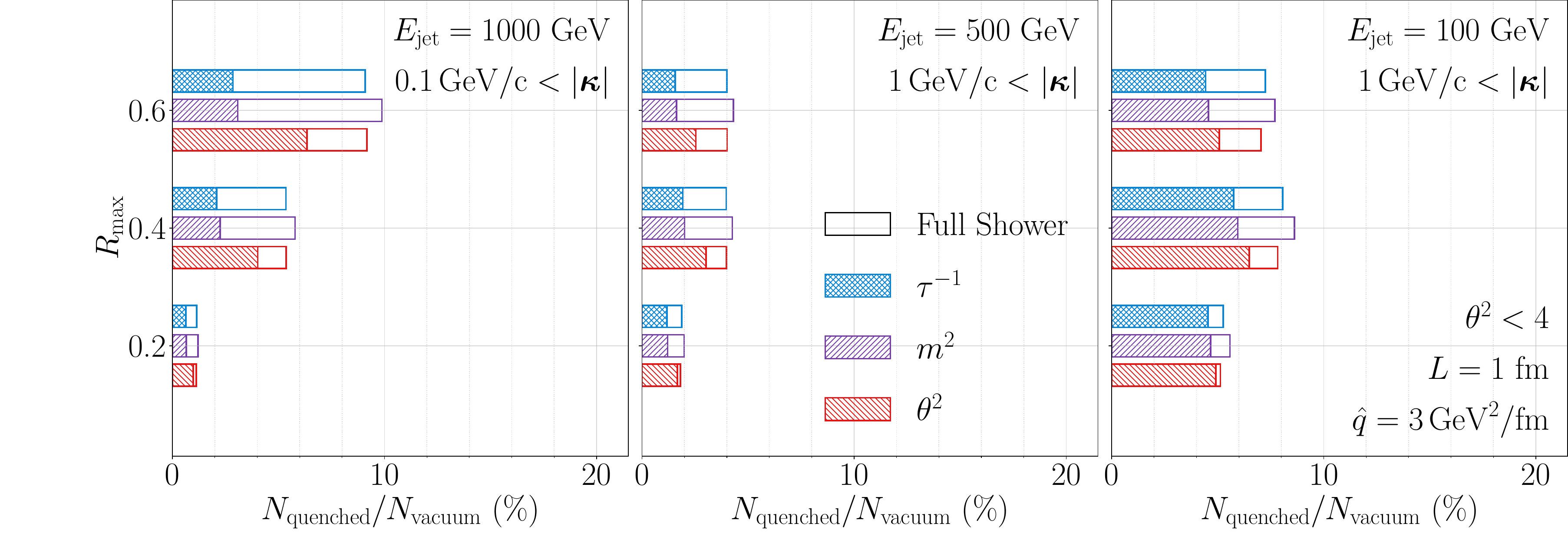}
\includegraphics[width=1.00\textwidth]{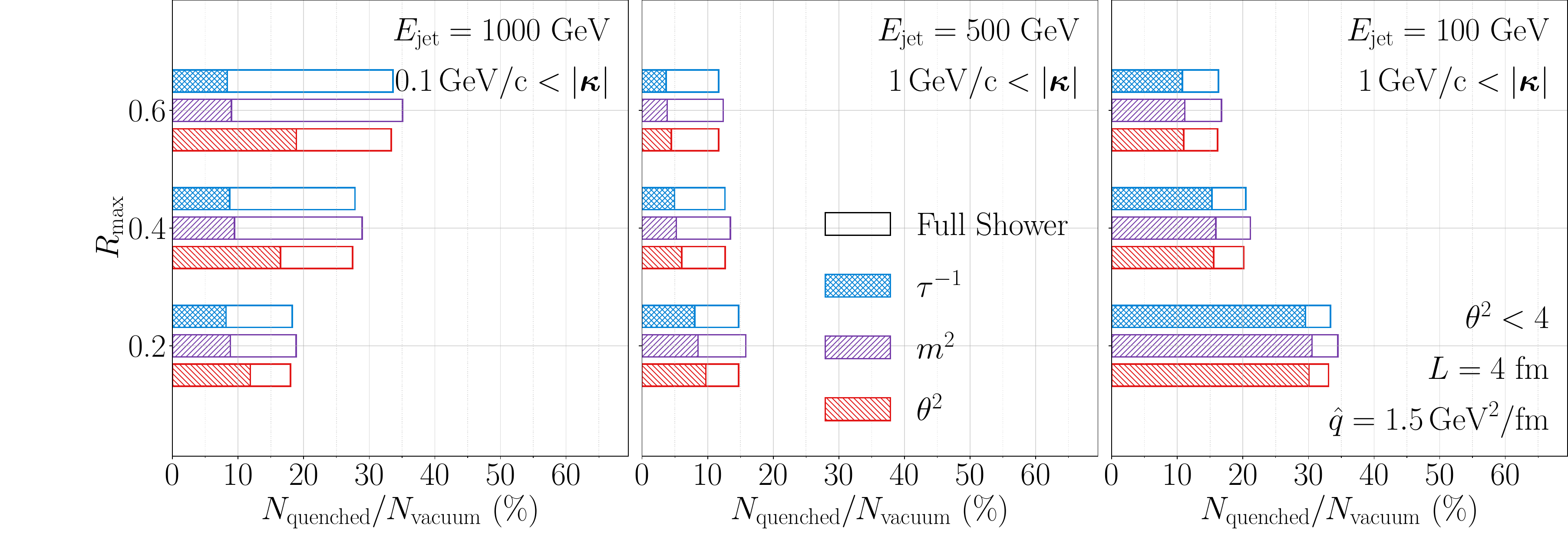}
\caption{
Percentage of events obeying the quenching condition defined by the first equation in eqs.~\eqref{eq:extra-quenching-conditions} for samples generated according to formation-time (blue), invariant-mass (purple), and angular (red) ordering prescriptions. Hatched rectangles correspond to the ``First Splitting'' mode, while empty rectangles to the ``Full Shower'' mode. %
Different panels correspond to differently chosen jet energies ($E_{\rm jet}$) and hadronisation scales for the vacuum samples.
Prior to applying the quenching condition, the samples were restricted to quark-branch splittings with $\angleVAR < R_{\rm max}$. The parameters chosen reflect either a short and dense \textbf{(top)}, or a long and dilute \textbf{(bottom)} medium.
}
\label{fig:quenching-weights-scalecomp}
\end{figure}

 We now examine the effects of varying the jet energy and hadronisation scales used to generate the vacuum samples. In figure~\ref{fig:quenching-weights-scalecomp}, we represent the effects of lowering the hadronisation scale to $\Lambda=\SI{0.1}{GeV/c}$ (left panel), and choosing two different jet energies, either $\SI{500}{GeV}$ (middle panel) or $\SI{100}{GeV}$ (right panel). The quenching probability corresponds to the first presented in the main text, see eq.~\eqref{eq:quench-prob}, and the medium parameters correspond to either the short-lived and dense (top panels), or the long-lived and dilute configuration (bottom panels).

We observe that differences between ordering prescriptions, as reflected in the quenching ratios, are amplified both by lowering the hadronisation scale and by considering a short lived medium. The former  (left panels) is due to the fact that extending the phase-space for emissions increases the magnitude of mismatches between ordering prescriptions. The latter arises because in a short-lived medium only splittings with short $\tform$  can interact with the medium, and these more susceptible to be sampled differently by different prescriptions. A similar line of reasoning explains why, with decreasing jet energies, medium effects become larger  but the relative differences between algorithms for the first splitting decreases. At smaller energies, fewer splittings occur, reducing the difference between first splitting and full shower. Splittings also tend to be wider and are therefore more likely to exceed the critical angle $\theta_{\rm dec}$, resulting in larger quenching ratios.

We note that for a lowered hadronisation condition the quenching ratios are monotonically increasing with the $R_{\rm max}$ parameter, unlike in the $\Lambda=\SI{1}{GeV/c}$ case discussed in the main text. This is because the reduced cut-off scale allows for narrower splittings, since $\theta > 4\Lambda/E$, which have parametrically longer decoherence times, $t_{\rm dec} \propto \theta^{-2/3} $. As a result, the dependence on $R_{\rm max}$ is purely explained by narrower vacuum-like splittings being more or less likely to occur inside the medium length $L$. On the other hand, when $\Lambda = \SI{1}{GeV/c}$, the $R_{\rm max}$ dependence for $\SI{100}{GeV}$ and $\SI{500}{GeV}$ showers follows the same trend as for the $\SI{1000}{GeV}$ case (c.f. left panels of figure~\ref{fig:quench-weights-modelcomp}),  although the magnitude of the quenching ratios is somewhat larger for lower energies and larger media.

\begin{figure}[h!]
\centering
\includegraphics[width=1.00\textwidth]{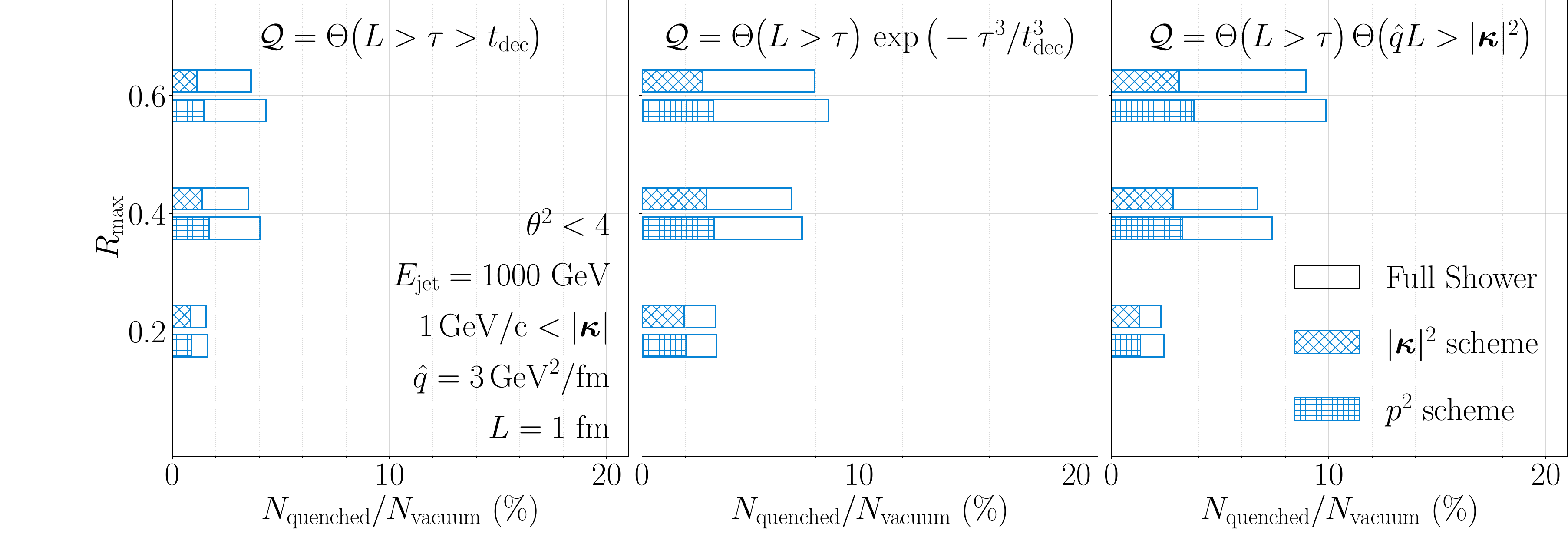}
\caption{
Percentage of events obeying the quenching condition in eqs.~\eqref{eq:extra-quenching-conditions} for samples generated according to formation-time ordering prescription in the momentum (blue crossed) and mass (blue squared) schemes.
Prior to applying the quenching condition, the samples were restricted to quark-branch splittings with $\angleVAR < R_{\rm max}$. The medium parameters used are $L=\SI{1}{fm}$, and $\hat{q}=\SI{3}{GeV^2/fm}$.
}
\label{fig:quenching-ratios-schemecomp}
\end{figure}

For completeness, we repeat our pseudo-quenching exercise, presented in figure~\ref{fig:quench-weights-modelcomp} for different evolution variables,  for different kinematic schemes. The results, shown in figure~\ref{fig:quenching-ratios-schemecomp}, reveal quantitative differences across all three models across and for all values of $R_{\rm max}$. Notably, the momentum scheme (blue crossed) exhibits lower quenching ratios than the mass scheme (blue squared). This result is somewhat counter-intuitive, as one might expect the time veto inherent to the $p^2$-scheme to effectively push emissions outside the medium. However, due to the bias towards lower transverse momentum in this scheme, the quenching ratios tend to be slightly larger, regardless of $R_{\rm max}$. 

Finally, these results show that the impact of the kinematic scheme on jet quenching is smaller than that of the ordering prescription  (cf. figure~\ref{fig:quench-weights-modelcomp}), especially when focusing on the collinear limit.

\bibliography{biblio}{}
\bibliographystyle{JHEP}

\end{document}